%% file: main.tex
\documentclass[sigconf]{acmart}   
\usepackage{xspace}
\usepackage{multirow}
\usepackage{pifont}
\usepackage{algorithmic}
\usepackage[ruled,linesnumbered,vlined]{algorithm2e}
\usepackage{tcolorbox}  
\usepackage{xcolor}
\usepackage{amsmath}
\usepackage{soul}
\usepackage{colortbl}

\usepackage{enumitem}
\usepackage{subfig}

\AtBeginDocument{%
  }

\setcopyright{acmlicensed}
\copyrightyear{2026}
\acmYear{2026}
\acmDOI{XXXXXXX.XXXXXXX}

\acmConference[ICSE 2026]{the 48th International Conference on Software Engineering}{12--18 April, 2026}{Rio de Janeiro, Brazil}

\begin{document}
\input{macro}

\title{Repair Ingredients Are All You Need: Improving Large Language Model-Based Program Repair via Repair Ingredients Search}

\author{Jiayi Zhang}
\authornote{Joint first authors, contributed equally with author ordering decided by \href{https://senseis.xmp.net/?Nigiri}{\textit{Nigiri}}.}
\affiliation{
  \institution{Nanyang Technological University}
  \country{Singapore}
}
\email{jzhang150@e.ntu.edu.sg}

\author{Kai Huang}
\authornotemark[1]
\affiliation{
  \institution{Technical University of Munich}
  \country{Germany}
}
\email{kai-kevin.huang@tum.de}

\author{Jian Zhang}
\authornote{Corresponding author.}
\affiliation{
  \institution{Nanyang Technological University}
  \country{Singapore}}
\email{jian_zhang@ntu.edu.sg}

\author{Yang Liu}
\affiliation{
  \institution{Nanyang Technological University}
  \country{Singapore}}
\email{yangliu@ntu.edu.sg}

\author{Chunyang Chen}
\affiliation{
  \institution{Technical University of Munich}
  \country{Germany}
}
\email{chun-yang.chen@tum.de}


\begin{abstract}
Automated Program Repair (APR) techniques aim to automatically fix buggy programs.
Among these, Large Language Model-based (LLM-based) approaches have shown great promise.
Recent advances demonstrate that directly leveraging LLMs can achieve leading results. However, these techniques remain suboptimal in generating contextually relevant and accurate patches, as they often overlook repair ingredients crucial for practical program repair. In this paper, we propose \TOOL, a novel framework that enables LLMs to autonomously search for repair ingredients throughout both the reasoning and solution phases of bug fixing. In the reasoning phase, \TOOL \space integrates static analysis tools to retrieve internal ingredients, such as variable definitions, to assist the LLM in root cause analysis when it encounters difficulty understanding the context. During the solution phase, when the LLM lacks experience in fixing specific bugs, \TOOL \space searches for external ingredients from historical bug fixes with similar bug patterns, leveraging both the buggy code and its root cause to guide the LLM in identifying appropriate repair actions, thereby increasing the likelihood of generating correct patches. Evaluations on two popular benchmarks (Defects4J V1.2 and V2.0) demonstrate the effectiveness of our approach over SOTA baselines. Notably, \TOOL \space fixes 146 bugs, which is 32 more than the baselines on Defects4J V1.2.
On Defects4J V2.0, \TOOL \space fixes 38 more bugs than the
SOTA. Importantly, when evaluating on the recent benchmarks that are free of data leakage risk, \TOOL \space also maintains the best performance. \chen{can be shortened.}
\kai{have removed some redundancies.}
\end{abstract}





\maketitle

\input{1-intro}

\input{2-motivation}
\input{3-approach}
\input{4-setup}
\input{5-evaluation}
\input{6-threats}
\input{7-relatedwork}
\input{8-conclusion}

\bibliographystyle{ACM-Reference-Format}
\bibliography{main}

\end{document}

%% file: macro.tex
\newcommand{\TOOL}{\textsc{ReinFix}}

\newcommand{\zj}[1]{\textcolor{blue}{\textbf{[#1]}}}

\newcommand{\kai}[1]{\textcolor{red}{[#1]}}

\newcommand{\jy}[1]{\textcolor{green}{\textbf{[#1]}}}

\newcommand{\chen}[1]{\textcolor{orange}{\textbf{[#1]}}}

\newcommand{\add}[1]{\textcolor{black}{#1}}

\renewcommand{\chen}[1]{} 
\renewcommand{\jy}[1]{} 
\renewcommand{\st}[1]{}
\renewcommand{\kai}[1]{}

%% file: 1-intro.tex
\section{Introduction}
\label{sec_intro}

\add{Automated Program Repair (APR) aims to automatically fix software bugs~\cite{APR_Survey_Martin,APR_Survey_Luca,APR_Survey_Legous,APR_Survey_Zhang,APR_Survey_Huang}.
Traditional APR techniques, including search-based~\cite{GenProg}, constraint-based~\cite{SemFix}, and template-based~\cite{TBar} methods, have made considerable strides in addressing common bug patterns. Template-based approaches, for example, have advanced in addressing common bugs but are restricted by the repair patterns to solve diverse bug types and complex repair scenarios~\cite{ChatRepair}. To overcome these limitations, learning-based methods, particularly Neural Machine Translation (NMT)-based APR approaches~\cite{NMT4APR_Tufano,SequenceR,DLFix,DEAR,CoCoNut,CURE,RewardRepair,Recoder,Tare,TRANSFER,TENURE}, frame bug fixing as a translation problem and learn the bug-fix domain knowledge from bug-fix pairs (BFPs)~\cite{NMT4APR_Tufano}, 
where buggy code is ``translated'' into correct code using a model trained on BFPs. 
}

Recently, the advent of Large Language Models (LLMs) has sparked a new wave of advancements in APR. Initially, many studies attempted to improve APR performance by fine-tuning LLMs, following similar strategies to NMT-based APR approaches~\cite{LLM4APR_Study_Huang,LLM4APR_Study_Jiang,LLM4APR_Study_Wu}\st{While this can be effective, these methods}\add{, which} remain constrained by limitations in training data. Therefore, more recent LLM-based approaches~\cite{ChatRepair,ThinkRepair,SRepair,ContrastRepair,LLM4APR_Study_Xia,GAMMA,AlphaRepair,RepairAgent,Repilot,GUIRepair,Agentless} leverage the vast amount of knowledge encoded in pre-trained models, such as ChatGPT and Codex, which have been trained on extensive datasets of open-source code. These models demonstrate impressive performance on APR tasks by generating possible patches directly. One notable example is ChatRepair~\cite{ChatRepair}, a ChatGPT-based APR approach that combines patch generation with instant feedback, enabling conversational-style repair interactions. 
Similarly, ThinkRepair \cite{ThinkRepair} uses Chain-of-Thought (CoT) prompting on ChatGPT to enhance bug reasoning during repair.

Despite their powerful bug-fixing capabilities, existing LLM-based APR works can struggle to produce contextually relevant and accurate patches when essential repair ingredients are absent~\cite{LLM4APR_Study_Huang_tse}.
\add{The repair ingredients is a concept built on the \textit{redundancy assumption}~\cite{RedundancyAssumptions_Study_Martinez}, which falls into two categories: internal and external~\cite{RepairIngredient_Study_Yang}.} Internal repair ingredients include key contextual elements (e.g., variables) essential for understanding or addressing the bug, while external ingredients consist of repair behaviors (e.g., change actions) from prior fixes that may guide effective patch generation. In fact, research has demonstrated the importance of this redundancy assumption (i.e., repair ingredients) in traditional APR approaches~\cite{RedundancyAssumptions_Study_Martinez}. On the one hand, redundant change actions from repair histories can often be distilled into fix patterns\jy{Should we unify it to `repair pattern` rather than `fix pattern`} (or repair templates), guiding template-based APR methods by providing actionable repair workflows~\cite{PAR}. On the other hand, redundant code fragments have played a vital role in search-based APR systems as donor code for creating patches~\cite{GenProg}. 
In the realm of LLM-based approaches, FitRepair~\cite{FitRepair} and RAP-Gen~\cite{RAP-Gen} are among the few representative works that utilize the redundancy assumption. FitRepair fine-tunes CodeT5 on the buggy project to learn internal code ingredients and prompts it to use these ingredients by searching relevant identifiers from the buggy project. Similarly, RAP-Gen retrieves external bug-fix pairs to augment the buggy input for the fine-tuning of CodeT5 as the patch generator.

However, existing strategies for using repair ingredients are primarily designed for small-scale LLMs and are not easily adaptable to advanced models.
In addition, previous methods were straightforward in searching and utilizing repair ingredients, which may have resulted in imprecise ingredient search and patch generation.
These issues stem from two key aspects: 
\textbf{1) Inaccurate Search for Repair Ingredients.} FitRepair collects internal code snippets exclusively from the buggy project for fine-tuning and searches for relevant identifiers through code similarity to enhance the prompted input, \st{While this enables the model to learn project-specific knowledge, it can also lead to the inclusion of}\add{which may include} irrelevant and redundant snippets\st{, introducing unnecessary noise}. Similarly, RAP-Gen retrieves external bug-fix pairs based solely on the similarity of code, neglecting high-level contextual information about the bug (e.g., its causes)\st{. This results in a disconnect between analyzing the current bug and retrieving appropriate fix actions.}\add{, leading to a disconnect between bug analysis and appropriate fix action retrieval.} \textbf{2) Limited Scalability of Repair Paradigms.} Both FitRepair and RAP-Gen utilize the fine-tuning paradigm based on CodeT5 to construct their repair workflows. \st{However, generalizing these approaches to mainstream large-scale LLMs (e.g., GPT-4) is challenging due to the high costs associated with fine-tuning LLMs, as well as the fact that some commercial models do not support fine-tuning. Consequently, the integration of repair ingredients to enhance repair capabilities in LLMs like ChatGPT family remains underexplored. In other words, the full potential of repair ingredients for program repair has yet to be unlocked in the era of LLMs.}
\add{Generalizing these approaches to mainstream LLMs (e.g., GPT-4) is challenging due to high cost of fine-tuning and it is not supported by some commercial models. Consequently, the integration of repair ingredients into LLMs to enhance repair capabilities remains underexplored, and the full potential of repair ingredients has yet to be unlocked in the era of LLMs.}

Inspired by recent advances in LLM-based agents for software engineering tasks~\cite{RepairAgent,FixAgent, Agent4SE_Survey_He,Agent4SE_Survey_Liu}, we recognize that these agents significantly enhance the versatility and expertise of LLMs by equipping them with the ability to perceive and utilize external resources and tools~\cite{Agent4SE_Survey_Liu}. We believe that enabling agents to autonomously invoke tools to search for repair ingredients presents a promising avenue for improvement.
To this end, we propose a novel framework, \TOOL~(\underline{Re}inforcement \underline{In}gredient \underline{Fix}er), which employs LLM-based agents to integrate internal and external repair ingredients into the reasoning and solution phases of patch generation, respectively. \TOOL~begins by reading the bug information, including buggy code and triggered test errors, and then initiates a planning step.
In the reasoning phase, where project-specific identifiers (e.g., unique Java classes) may be unfamiliar to the LLM, \TOOL~equips it with a toolkit capable of searching for definitions and details of those elements as internal repair ingredients via dependency analysis. This setup allows the LLM to invoke the tools by passing unknown identifiers as arguments. The additional information supports more accurate root-cause analysis, aiding in devising a valid solution for the bug.
In the solution phase, \TOOL~ guides the LLM to analyze root causes from historical bug fixes, which are vectorized using an embedding model and stored in a vector database. Instead of solely considering code similarity as \st{previous work~\cite{RAP-Gen}}\add{RAP-Gen} did, we design a retrieval tool that searches for a \st{historical}\add{external} bug-fix pair with both similar bugs and root causes\st{. These external bug-fix pairs serve as repair patterns} \add{as repair patterns}, enabling the LLM to generate more accurate and contextually relevant suggestions for the final generation of patches. 
To coordinate these processes, we implement our approach on the basis of the Reasoning and Acting framework~\cite{ReAct}, empowering the LLM to autonomously plan, analyze, and invoke the appropriate repair ingredients search tools as needed. This means that all tools are provided as optional. When the bug is straightforward or the LLM is confident based on prior knowledge, it may bypass certain tools. This flexible, integrated approach strengthens both the reasoning and solution phases, which eventually enhances the quality of generated patches.
In summary, this paper makes the following contributions:
\begin{itemize}[leftmargin=0.3cm]
    \item \textbf{Dimension.} 
    This paper opens up a new dimension for LLM4APR research to incorporate repair ingredients to enhance the model's repair capabilities in the era of LLMs.
    Unlike recent LLM4APR work, which either uses only donor code~\cite{FitRepair} (i.e., internal repair ingredients) or only change actions~\cite{RAP-Gen} (i.e., external repair ingredients), our work effectively integrates both internal and external repair ingredients with LLMs to boost APR performance and alleviate the limitations of previous works. 
    Our work showcases, for the first time, a promising path toward integrating LLM-based agents with repair ingredients.

    \item \textbf{Extensive Study.}
    We performed extensive evaluations on Defects4J V1.2 and Defects4J V2.0. The results demonstrate that \TOOL~outperforms previous APR tools and further boosts the repair capabilities of LLMs. For instance, with GPT-3.5 as the foundational model, \TOOL~successfully repairs 118 and 123 bugs in Defects4J V1.2 and V2.0, respectively, surpassing the top baseline APR tools by 4 and 16 bugs. Similarly, utilizing the more substantial GPT-4 model, \TOOL~fixes 146 and 145 bugs in Defects4J V1.2 and V2.0, respectively, outperforming the best baselines by 32 and 38 bugs. 
    Notably, \TOOL's application with GPT-4 fixed 61 (146-85) more bugs than the foundation model in our ablation study, marking a 71.76\% improvement. 

    \item \textbf{Open Science.}
    We have released \TOOL~framework with two novel repair ingredients searching tools. Our repair ingredients retrieval scheme can be flexibly integrated into many LLM-based APR tools to further advance the APR community. We also encourage future researchers to leverage our approach to develop more powerful APR tools.
    Our source code and data are available at: \url{https://sites.google.com/view/repairingredients}.

\end{itemize}

%% file: 2-motivation.tex
\section{Motivation}
\label{sec_motivaton}

\add{
Despite the potential of LLMs in understanding code, they face challenges of lacking the ability to gather contextual clues and prior knowledge of similar fixes (i,e., internal and external repair ingredients) independently to generate effective patches. 
In this section, we explore how these repair ingredients are critical to debugging and repair, using examples to show how missing ingredients can hinder effective patch generation.
}


\st{When developers debug and fix software, they often rely on contextual clues and prior knowledge of similar fixes. Repair ingredients, which consist of essential information and patterns that aid in crafting effective patches, are crucial to this process. These repair ingredients include internal details specific to the current project (like variable names or unique classes) and external patterns from past fixes that guide the solution.}

\st{Despite the potential of LLMs in understanding code, they face challenges that developers do not encounter as often: they lack the ability to gather these internal and external repair ingredients independently. Below, we explore how these ingredients are critical to debugging and repair, using examples to show how missing ingredients can hinder effective patch generation.}


\subsection{Example 1: The Role of Internal Ingredients}
\begin{figure}[!t]
\centering
\begin{center}
\subfloat[The human patch of Closure-14.]{
		\includegraphics[width=1\linewidth]{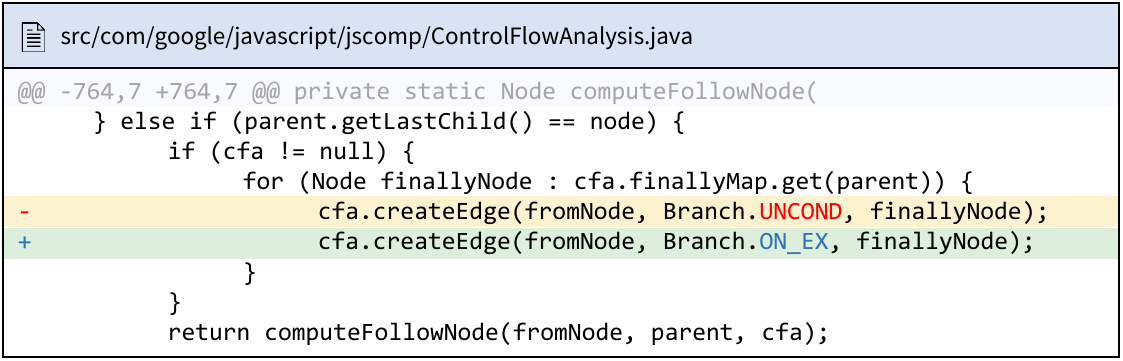}}
\end{center}
\begin{center}
\subfloat[Internal repair ingredients for Closure-14.]{
		\includegraphics[width=1\linewidth]{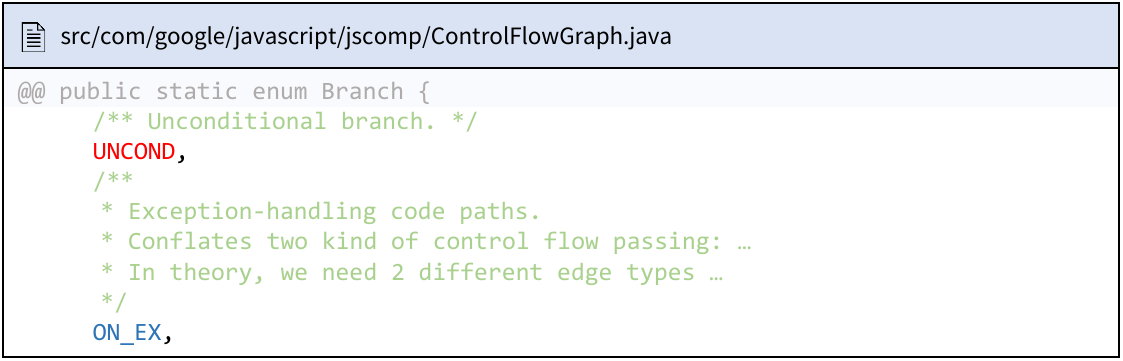}}
\end{center}
\caption{A motivation example of Closure-14.}
\label{fig:closure-14}
\end{figure}

Figure \ref{fig:closure-14}-(a) illustrates the Closure-14 bug from Defects4J, which occurs within Google’s Closure Compiler project in the \texttt{ControlFlow}\-\texttt{Analysis.java} file, specifically within the \texttt{computeFollowNode} method. This method is responsible for calculating the next node in control flow after a given node in an Abstract Syntax Tree (AST) is executed. However, the buggy code incorrectly uses \texttt{Branch.UNCOND}, an unconditional branch setting, to create edges for final nodes. This choice leads to an incorrect representation of the control flow because it fails to capture the exception-handling path correctly.
The human patch corrects this by replacing \texttt{UNCOND} with \texttt{ON\_EX}, ensuring the control flow graph accurately reflects that the ``finally'' block executes during exception propagation.

For a developer familiar with the Closure project, \texttt{Branch} and its attributes like \texttt{UNCOND} and \texttt{ON\_EX} are meaningful, since they indicate specific control flow behaviors. However, without access to this internal information, the LLM faces challenges in understanding why \texttt{UNCOND} is incorrect in this context or why replacing it with \texttt{ON\_EX} is effective. If the LLM only has access to the buggy function itself, synthesizing the correct patch becomes difficult due to the lack of contextual understanding about these parameters and their correct application.
To solve this bug, essential donor code containing internal repair ingredients is needed. In this example, searching through the project reveals that the definition of \texttt{Branch} can be found in a non-buggy file, \texttt{ControlFlowGraph.java}. As illustrated in Figure~\ref{fig:closure-14}-(b), by collecting and analyzing these internal repair ingredients, both within the buggy file and across other files in the project, the LLM can build a more comprehensive understanding of the needed context, ultimately leading to an effective fix.

\subsection{Example 2: The Role of External Ingredients}

\begin{figure}[!t]
\centering
\begin{center}
\subfloat[The human patch of Closure-51.]{
		\includegraphics[width=1\linewidth]{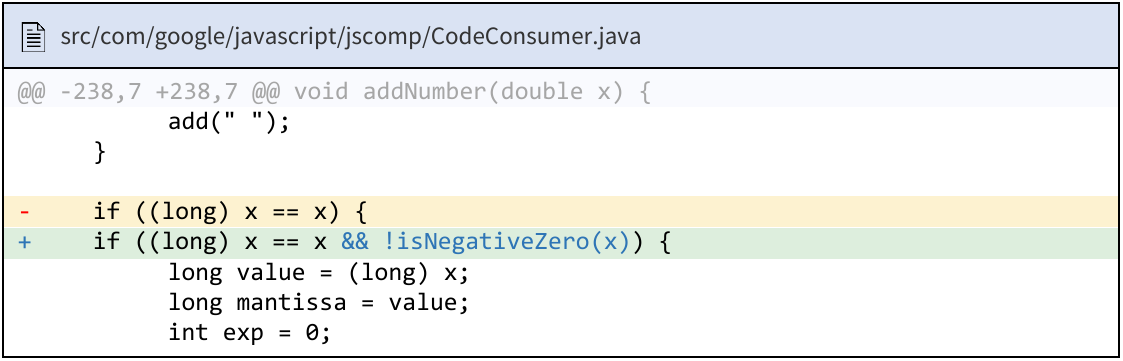}}
\end{center}
\begin{center}
\subfloat[External repair ingredients for Closure-51.]{
		\includegraphics[width=1\linewidth]{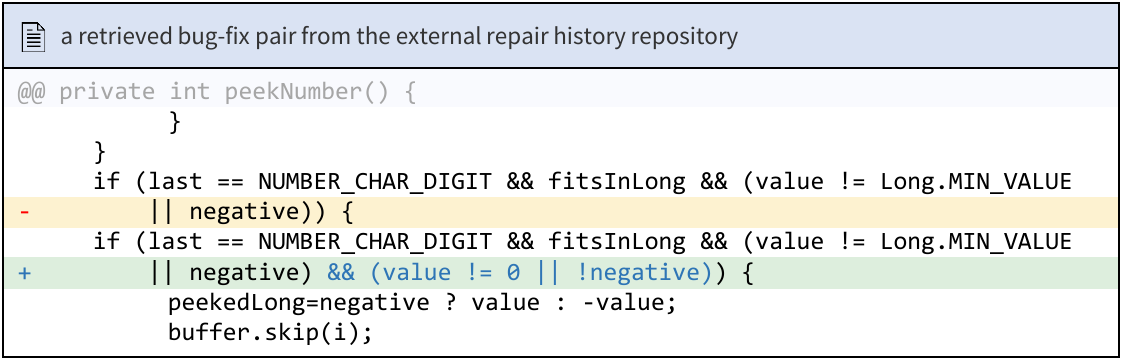}}
\end{center}
\caption{A motivation example of Closure-51.}
\label{fig:closure-51}
\end{figure}

Figure~\ref{fig:closure-51}-(a) illustrates another bug, Closure-51. This bug reveals an incomplete handling of a floating-point edge case in which the code uses \texttt{(long) x == x} to verify whether a floating-point number \textit{x} can be converted to a long integer without loss of precision. This condition works as intended for most values except \textit{x} is equal to negative zero,
allowing this specific value to \add{proceed, and}\st{pass through to other parts of the code. This can} lead to unexpected behaviors and potential errors in subsequent code execution due to the peculiar properties of negative zero in floating-point arithmetic. The human-created patch addresses this by adding an additional condition, \texttt{!isNegativeZero(x)},
\add{explicitly preventing negative zero from incorrectly passing the conversion condition.} From a developer's perspective, the need for this patch is clear: without explicitly handling negative zero, the code risks unpredictable behaviors that can complicate program correctness.

However, this type of subtle bug presents a challenge for LLM-based approaches in APR. Despite the possibility that LLMs may have encountered similar edge cases in training, reliably generating the correct patch requires specialized domain knowledge about floating-point behavior and precision conversion. Smaller LLMs, in particular, might lack this depth of understanding. In this scenario, drawing on external repair history, which is knowledge about prior similar bugs and patches, could significantly enhance the LLM's ability to generate accurate fixes.
For example, as illustrated in Figure~\ref{fig:closure-51}-(b), repair history records containing analogous bug-fix pairs provide a valuable resource for understanding and addressing the current bug. Here, a repair history entry shows a bug fix with a similar structure, where a negative zero check is added to address a floating-point precision issue. 
\add{These external repair ingredients offer concrete repair behaviors, guiding LLMs to craft effective patches for bugs by referencing prior solutions of similar issues.}

\st{In summary, these examples highlight the essential role of repair ingredients in program repair. Drawing on insights from developer practice, these ingredients improve \add{bug comprehension} and \add{patch} generation\st{phases of patch creation}, enabling more precise and contextually relevant fixes. In the following section, we demonstrate how this knowledge can be effectively integrated into an LLM-based APR framework to systematically strengthen the patch repair process.}

%% file: 3-approach.tex
\section{Approach}\label{sec_approach}

\begin{figure*}[]
    \centering
    \vspace{-0.1in}
    {\includegraphics[width=1.0\linewidth]{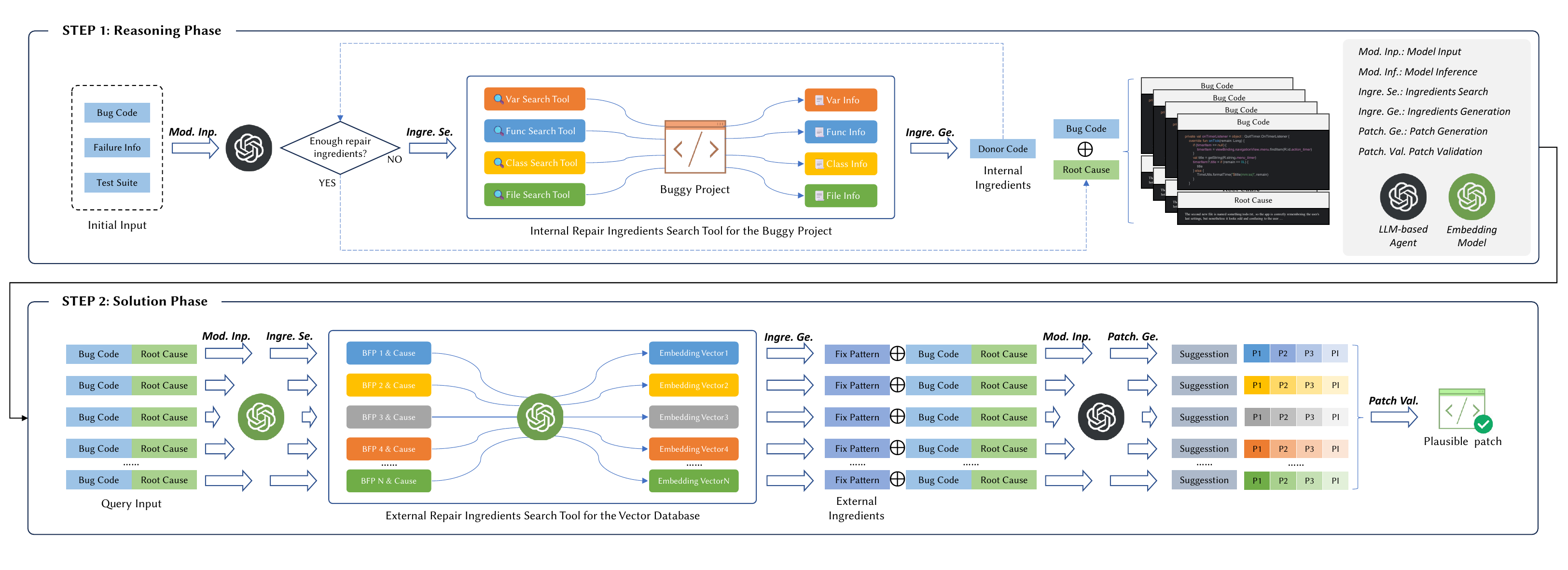}}
    \vspace{-0.4in}
    \caption{The workflow of \TOOL.} 
    \label{fig:overview}
    \vspace{-0.1in}
\end{figure*}

In this section, we introduce \TOOL, which was designed to enhance LLM-based APR by incorporating both internal and external repair ingredients. As shown in Figure~\ref{fig:overview}, \TOOL \space operates in two primary phases: reasoning and solution. 
In the reasoning phase, \TOOL \space supplies internal ingredients, which are derived through a toolkit that conducts dependency analysis to identify and retrieve relevant contextual code elements (e.g., definitions and usages of key variables or methods) within the project. This approach enables the LLM to gain a deeper understanding of project-specific dependencies and key identifiers, which aids in root cause analysis.
In the solution phase, \TOOL~enhances the patch generation by leveraging external ingredients. Specifically,
\add{the LLM retrieves similar examples from a vector database of bug-fix pairs based on the current bug and its root cause, using them as a reference for patch generation.} The generated patches then undergo the validation process to output the plausible one which passes all test cases.

\subsection{Internal Ingredient Search}
\label{sec_approach:donor}

In the reasoning phase, \textit{internal ingredients} refer to critical code elements that aid in the bug cause comprehension. These include variables, methods, classes, and file dependencies essential for understanding the bug context and crafting an accurate patch. Existing works, such as \textit{FitRepair}~\cite{FitRepair}, rely on embedding relevant identifiers or variable names by performing similarity-based searches. While this approach can be effective, it has limitations when handling complex, cross-functional dependencies or cases requiring deeper contextual understanding. To address these challenges, \TOOL \space integrates a structured \textit{dependency analysis} to identify and retrieve internal ingredients across various program structures.

Rather than embedding only variables or identifiers, we leverage \textit{Joern}~\cite{joern_tool} to analyze the codebase through a \textit{Code Property Graph (CPG)}. The CPG enables sophisticated queries across the code, helping gather multi-level dependencies for bug understanding. \TOOL \space uses Joern-based queries across four levels of analysis including \textit{Variable}, \textit{Method}, \textit{Class}, and \textit{File}, as shown in Table~\ref{tab:joern-tools}.

\input{tables/Joern_Tools}

\begin{itemize}[leftmargin=0.3cm]
    \item \textbf{Variable-Level Analysis}: 
    Variable-level tools focus on tracking variable definitions, usages, and types within specific files, providing precise information about how variables are defined and interact with other components. For example, the \texttt{identify\_variable} tool enables targeted retrieval of variable definitions:\jy{should reformat these formula, is it ok to reform like this:?}
    \begin{multline*}
    \text{identify\_variable}(\text{varName}, \text{fileName}) = \\
    \{ v \mid v \in C_{\text{fileName}}, \text{is\_var}(v) \land \text{matches}(v, \text{varName}) \}
    \end{multline*}
    where \(C_{\text{fileName}}\) denotes all code elements in the file, returning matching variables by name. Additional tools like \texttt{find\_variable}\-\texttt{\_assignments} and \texttt{track\_variable\_data}\-\texttt{flow} provide details on value assignments and data flow through variables, offering the LLM insights into how variables influence the program state.

    \item \textbf{Method-Level Analysis}: 
    The \texttt{trace\_method\_usage} tool allows the retrieval of all method invocation locations, regardless of where they appear in the project:
    \begin{multline*}
    \text{trace\_method\_usage}(\text{methodName}) = \\ 
    \{ (f, l) \mid \text{invokes}(C_f, \text{methodName}) \text{ at location } l \}
    \end{multline*}
    
    This query lists each location where \texttt{methodName} is invoked, providing control flow context essential for understanding program behavior. The \texttt{analyze\_method\_details} tool additionally extracts detailed control structures within the method, such as loops and exception handling, providing a richer basis for understanding how specific actions influence the program state.

    \item \textbf{Class-Level Analysis}: 
    Classes represent higher-level structures that encapsulate variables and methods, and understanding class definitions can be vital in cases where multiple elements interact through shared methods or attributes. The \texttt{identify\_class} tool can retrieve class definitions, including attributes and methods:
    \begin{multline*}
    \text{identify\_class}(\text{className}) = \\ 
    \{ c \mid c \in C_{\text{fileName}}, \text{is\_class}(c) \land \text{matches}(c, \text{className}) \}
    \end{multline*}
    This query retrieves the definition of \texttt{className}, giving the LLM insights into the encapsulated methods and properties that may influence the bug context.   Similarly, the \texttt{find\_class\_loc} tool enables the LLM to pinpoint where the class is defined, helping to trace potential interactions and dependencies that are critical to understanding complex bug scenarios.

    \item \textbf{File-Level Analysis}: 
    Since projects often span multiple files, file-level tools allow us to track dependencies, such as imports and inter-file references, that are crucial for understanding how components interact across files. For example, the \texttt{get\_imports} tool gathers all imports in a file:
    \[
    \text{get\_imports}(\text{fileName}) = \{ i \mid i \in \text{imports}(C_{\text{fileName}}) \}
    \]
    This information is essential when dependencies from external libraries or other modules impact the buggy code’s behavior.
\end{itemize}

\begin{figure}[t]
    \centering        
    {\includegraphics[width=1\linewidth]{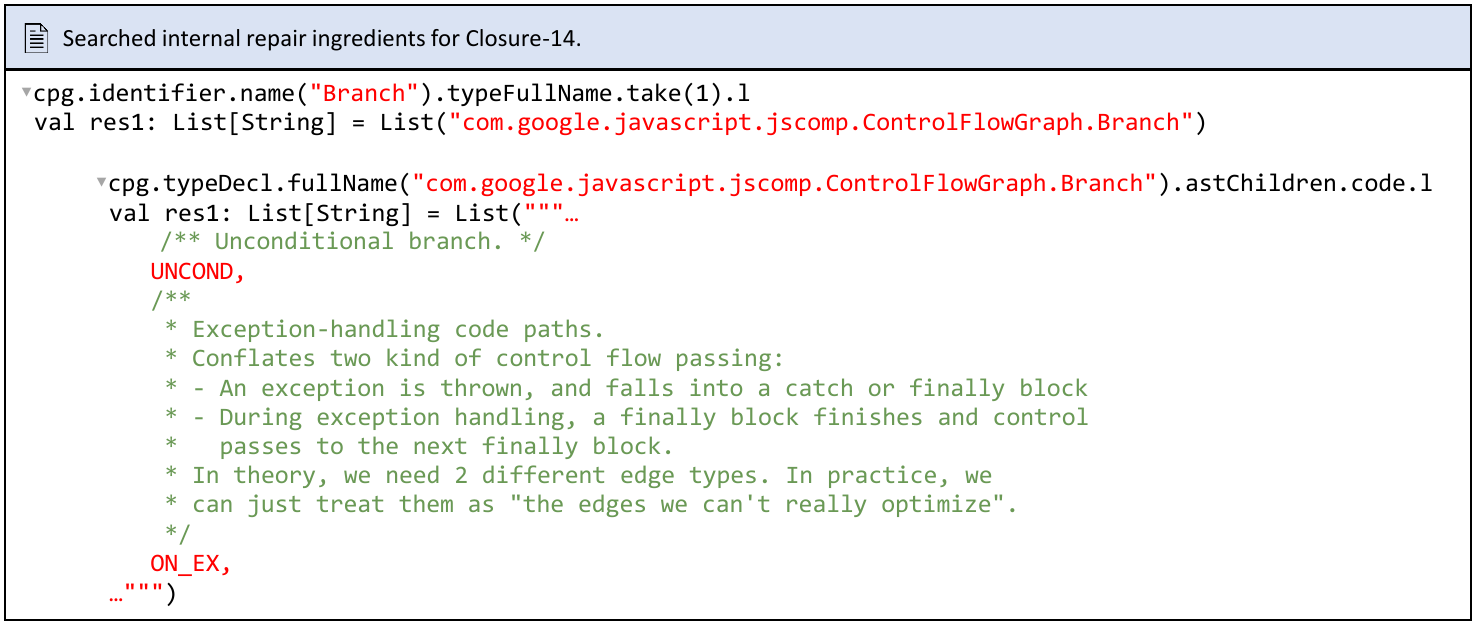}}
    \vspace{-0.2in}
    \caption{Searched class-level dependencies for Closure-14.}
    \label{fig:Closure_14_donor_code}
    \vspace{-0.2in}
\end{figure}

\TOOL \space extracts comprehensive internal ingredients 
from the codebase\st{. This depth of context enables}\add{, enabling} the LLM to go beyond simple identifier matching and to trace complex dependencies that often hold the key to uncovering a bug's root cause. By thoroughly analyzing variable definitions, method interactions, and control flows, \TOOL \space equips the LLM with a holistic understanding of how different components interact, which is crucial for identifying where and why errors arise.
For instance, in \textit{Closure-14} (Figure~\ref{fig:closure-14}), the bug stems from an incomplete representation of control flow in the \texttt{computeFollowNode} method, where the code mistakenly uses \texttt{Branch.UNCOND} instead of \texttt{ON\_EX}. 
As shown in Figure~\ref{fig:Closure_14_donor_code},
by leveraging the \texttt{find\_class\_loc} tool and \texttt{identify\_class tool}, the LLM can retrieve information on where class \texttt{Branch} locates and search for the definition code of this class, including comments of the attributes. It is clear that \texttt{ON\_EX} from other parts of the project, although this identifier is absent from the buggy code. This allows the LLM to identify that \texttt{ON\_EX} is the correct control flow branch to use in exception propagation, illuminating the underlying \add{root cause}. \st{This approach facilitates a form of root cause analysis, where the LLM can trace why the existing implementation fails to represent the control flow accurately.}


\subsection{External Ingredient Search}
\label{sec_approach:pattern}

In the solution phase, \TOOL \space incorporates external repair ingredients derived from a large corpus of historical bug-fix pairs. These external ingredients, consisting of buggy code and their corresponding fixes, are essential to provide the LLM with relevant repair knowledge. Using this historical corpus, \TOOL~can offer insights into common bug patterns, allowing the LLM to reference and adapt known solutions to current bugs. This is crucial as many software bugs exhibit recurring patterns, and leveraging previous repairs can greatly improve the patch efficiency and accuracy~\cite{TBar,TRANSFER,TENURE,NTR,RAP-Gen}.
However, directly adopting the standard Retrieval-Augmented Generation (RAG), such as RAP-Gen~\cite{RAP-Gen}, \st{which retrieves similar buggy code and its fixes based on code similarity alone, is not sufficient for accurate bug fixing. In RAP-Gen,} which focuses on identifying similar buggy code, then retrieving the corresponding fix. While this is effective in some cases, it neglects the deeper context-specifically, the root cause of the bug. Bugs of similar symptoms might have different underlying causes, and relying solely on code similarity does not guarantee the identification of an appropriate fix. \st{For example, two bugs exhibiting similar code patterns may arise from different logical or environmental issues. Therefore, simply matching buggy code without considering the root cause may lead to irrelevant or incorrect repair suggestions.}

\begin{figure}[]
    \centering
    {\includegraphics[width=1.0\linewidth]{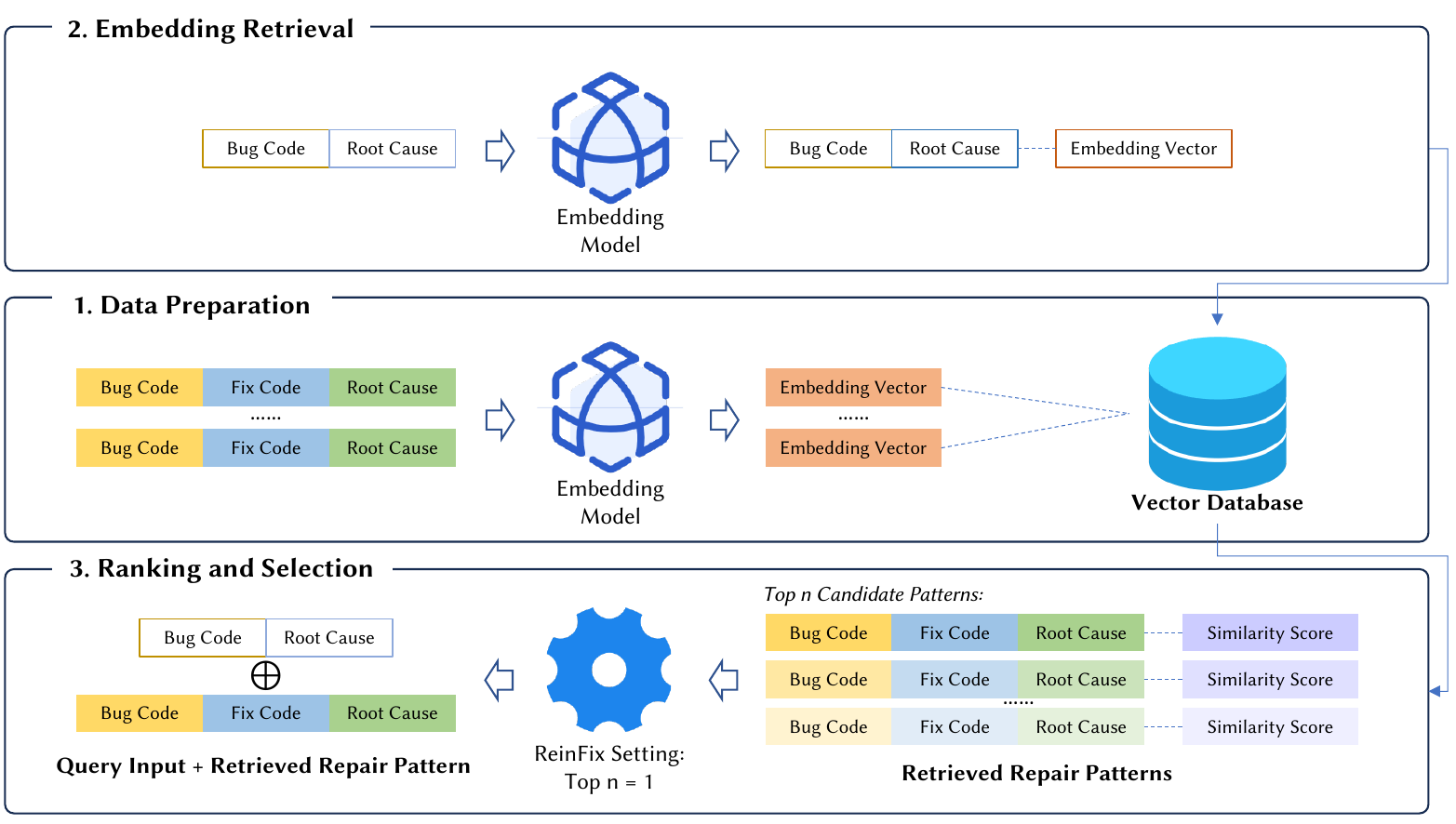}}
    \vspace{-0.2in}
    \caption{The workflow of retrieving external repair patterns through the enhanced RAG paradigm. 
    }
    \vspace{-0.2in}
    \label{fig:retrieval}
\end{figure}

To overcome these limitations, we enhance the corpus of BFPs by generating concise causes via prompting an LLM (e.g., GPT-4). For each buggy code and patch pair, forming structured triads of (buggy\_code, fix\_code, root\_cause). This augments traditional RAG by integrating both buggy code and root cause into the retrieval process, ensuring retrieved repair patterns are contextually and semantically relevant. The retrieval process, shown in Figure~\ref{fig:retrieval}, comprises three stages: Data Preparation, Embedding Retrieval, and Ranking and Selection.

\begin{itemize}[leftmargin=0.3cm]
    \item \textbf{Data Preparation.} First, we collect a large amount of historical bug-fix pairs from open-source code repositories and use LLMs (e.g., we use GPT-4o in our experiment) to automatically label the root cause for each BFP. Then we load each (\textit{buggy\_code}, \textit{fix\_code}, \textit{root\_cause}) triad from the corpus. For efficient retrieval, we preprocess the corpus by embedding each buggy code and its root cause. Let each buggy code \( c_i \) and its corresponding root cause \( r_i \) be part of a pair \( (c_i, r_i) \) from the corpus. We concatenate \( c_i \) and \( r_i \) into a single input string and transform it into a fixed-length embedding vector \( e_i \) using a pre-trained text embedding model (e.g., \textit{text-embedding-3-large}~\cite{Vector_Embeddings}):
\begin{equation}
    e_i = f_{\text{embed}}(c_i \oplus r_i)
\end{equation}
where \( \oplus \) denotes the concatenation operation. These embeddings \( e_i \) are stored in a vector database alongside the original corpus, resulting in a structure of (\textit{buggy\_code}, \textit{fix\_code}, \textit{root\_cause}, \textit{embedding\_value}). This embedded structure enables fast and accurate similarity comparisons for retrieving relevant repair patterns during the retrieval phase.

    \item \textbf{Embedding Retrieval.} Upon receiving a query in the form of \texttt{"<buggy code, root cause>"}, the query is also transformed into an embedding vector \( q \), which combines the buggy code and the root cause from the reasoning phase:
    \begin{equation}
        q = f_{\text{embed}}(\textit{query\_code} \oplus \textit{root\_cause})
    \end{equation}
    This query embedding is then compared to the embeddings in the precomputed corpus using cosine similarity:
    \begin{equation}
        \text{similarity}(q, e_{embedding}) = \frac{q \cdot e_{embedding}}{\|q\| \|e_{embedding}\|}
    \end{equation}
    The goal is to find historical bug-fix pairs that share similar structural and semantic characteristics with the query.

    \item \textbf{Ranking and Selection.} The top \( n \) results are retrieved based on similarity scores. For each retrieved entry, the corresponding buggy code, fix code, and root cause are returned to the LLM \jy{when the similarity score is larger than a certain threshold to deduct irrelevant pattern}along with the similarity score. These retrieved patterns serve as potential repair behaviors that the LLM can use to generate a correct patch for the current bug. Formally, let \( R \) denote the set of top \( n \) entries:
    \begin{equation}
        R = \{ e \in \text{corpus} \mid \text{similarity}(q, e_{embedding}) \geq \text{threshold} \}
    \end{equation}
    The LLM then uses these repair behaviors to inform its patch generation, ensuring that the proposed fix is grounded in prior successful repairs.

\end{itemize}

\begin{figure}[t]
    \centering
    {\includegraphics[width=1\linewidth]{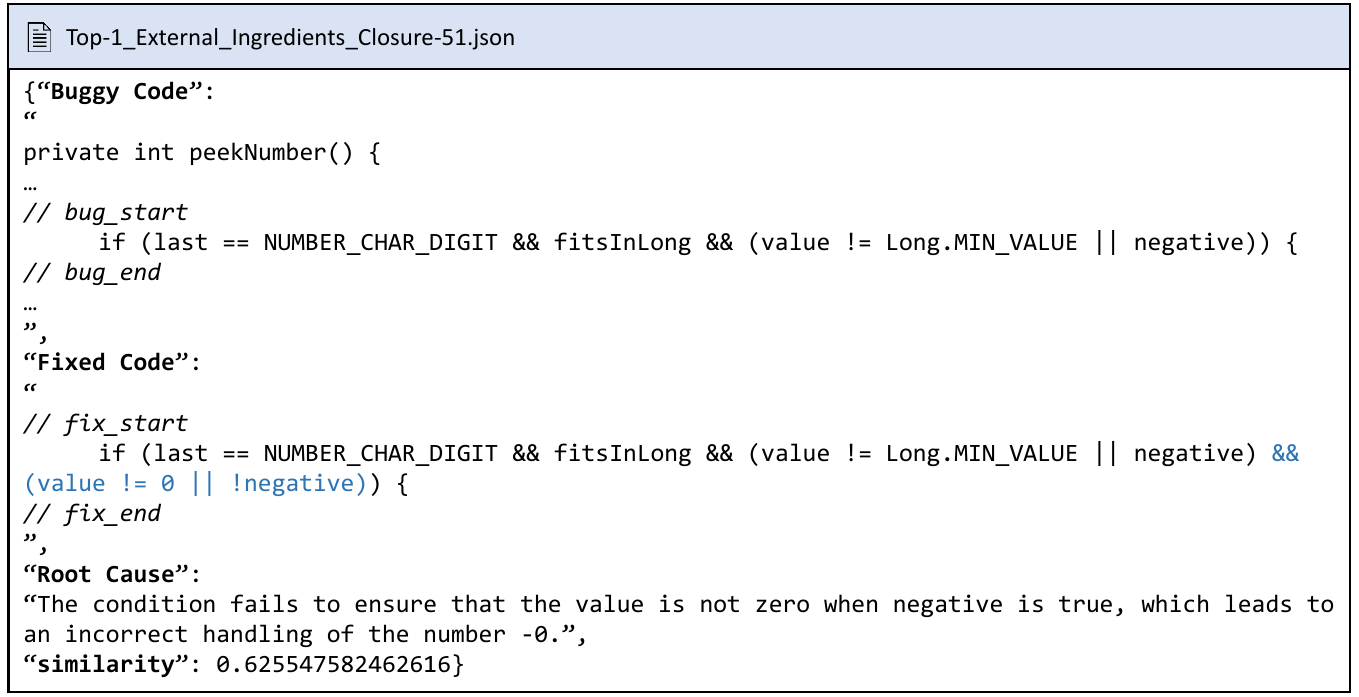}}
    \vspace{-0.2in}
    \caption{Retrieved repair pattern for \textit{Closure-51}.}
    \label{fig:repair_pattern}
    \vspace{-0.2in}
\end{figure}

In the \textit{Closure-51} (Figure~\ref{fig:closure-51}), \TOOL \space first queries the bug corpus using the buggy code and a preliminary root cause. The retrieved repair pattern, shown in Figure~\ref{fig:repair_pattern}, matches the root cause related to negative zero and suggests an appropriate fix. By referring to this repair pattern, \TOOL \space is also able to generate a patch that addresses the underlying issue. This process demonstrates how combining both buggy code and root cause information improves the relevance of the retrieved repair patterns.

\subsection{Overall Workflow}
\label{sec_approach:workflow}

\TOOL \space was built on the ReAct (Reasoning and Action) framework, completing  adaptive and context-aware program repair through \textit{Thought-Action-Observation} cycles.
In the \textbf{Thought Cycle}, the agent identifies which ingredients are essential for the bug comprehension. In the \textbf{Action Cycle}, where the agent invokes relevant tools based on the Thought Cycle's reasoning, conducting contextual code element analysis and external ingredient retrieval. Then, the agent interprets these tools' outputs to integrate new knowledge into its repair context during \textbf{Observation Cycle}, deciding on whether to use this information to refine its search or, if sufficient, proceed toward patch generation.
Through this iterative process, the agent reasons about necessary steps, selectively calls tools, and interprets tool outputs to make informed decisions.
This structured process adaptively searches internal codebase and external bug-fix patterns, efficiently synthesizing insights for patch generation.

\st{
Specifically, the ReAct Agent operates through \textit{Thought-Action-Observation} cycles, where each cycle incrementally advances \TOOL’s understanding and solution development. Through this iterative process, the agent reasons about necessary steps, selectively calls tools to gather data, and interprets tool outputs to make informed decisions. This structured progression allows \TOOL \space to adaptively search for ingredients across the codebase (internal search) and in historical bug-fix patterns (external search), efficiently synthesizing insights for patch generation.

\begin{itemize}[leftmargin=0.3cm]
    \item \textbf{Thought Cycle}: Each cycle begins with the ReAct Agent analyzing the current repair context. In this reasoning step, the agent identifies which ingredients are essential for understanding the bug or producing a repair. For example, if specific variable dependencies need clarification, the agent might plan to call variable-level analysis tools. Alternatively, if historical patterns could clarify the potential repair behavior, the agent would consider searching the corpus of previous bug-fix pairs.
    
    \item \textbf{Action Cycle}: Based on the reasoning from the Thought cycle, the agent selects and calls appropriate tools. The toolkit includes specialized utilities for contextual code element analysis (e.g., dependency tracing, variable data flow, and method usage analysis, as discussed in Section~\ref{sec_approach:donor}) and external ingredient retrieval using Retrieval-Augmented Generation techniques (detailed in Section~\ref{sec_approach:pattern}). This flexibility allows the agent to balance contextual code elements and external insights, which collectively inform the repair process.
    
    \item \textbf{Observation Cycle}: After the tools have run, the agent interprets their outputs to integrate new knowledge into its repair context. If a new root cause hypothesis or relevant contextual elements emerge, the agent can use this information to refine its search or, if sufficient, proceed toward patch generation.
\end{itemize}
}

Algorithm~\ref{alg:RepairIngredientFixer} outlines the overall workflow of \TOOL. \textit{1) Reasoning Phase}: This phase involves the agent’s primary exploration of root causes and gathering relevant repair ingredients (lines 3-5). If the initial root cause is inconclusive, the agent initiates an iterative search by invoking tools to collect contextual elements for reasoning (lines 6-22), constantly refining its analysis until the information is sufficient for analyzing the root cause.
\textit{2) Solution Phase}: The agent enters to this phase once the necessary repair ingredients are obtained. Here, it synthesizes a repair plan by integrating the internal and external insights (lines 23-25). The agent assembles the root cause, buggy code, and relevant repair patterns to generate repair suggestions. These suggestions form the basis of candidate patches, which are then validated against test cases to ensure reliability and relevance (lines 26-27).

\begin{algorithm}[t]
\footnotesize
\caption{\TOOL}
\label{alg:RepairIngredientFixer}
\DontPrintSemicolon
\SetKwProg{Framework}{Framework}{:}{}
\SetKwInOut{Input}{Input}
\SetKwInOut{Output}{Output}
\SetKwProg{Start}{Start}{:}{}
\Framework{\TOOL}{
    \Input{bugCode (buggy function code), failureInfo (failing test info), testSuite (test suite)}
    \Output{pPatch (plausible patch)}
    \Start{Reasoning Phase}{ 
        rootCause $\leftarrow$ CauseAnalyzewithoutDonor(bugCode, failureInfo, testSuite)\;
        \If{rootCause is not NONE}{
            \Return rootCause\;
        }
        \Else{
            Open CPG Project\;
            donorCode, varInfo, funcInfo, classInfo, fileInfo $\leftarrow \varnothing, \varnothing, \varnothing, \varnothing, \varnothing$\;
            isEnoughDonor $\leftarrow$ False\;
            analyze\_variable\_tool, analyze\_method\_tool, analyze\_class\_tool, analyze\_file\_tool $\leftarrow$ toolkit\;
            
            \While{isEnoughDonor is False}{
                varName, funcName, className, fileName $\leftarrow$ ingredientsSelection(bugCode, donorCode)\;
                varInfo $\leftarrow$ analyze\_variable\_tool(varName)\;
                funcInfo $\leftarrow$ analyze\_method\_tool(funcName)\;
                classInfo $\leftarrow$ analyze\_class\_tool(className)\;
                fileInfo $\leftarrow$ analyze\_file\_tool(fileName)\;
                donorCode $\leftarrow$ donorCode $\cup$ \{varInfo, funcInfo, classInfo, fileInfo\}\;
                rootCause $\leftarrow$ causeanalyzewithDonor(bugCode, donorCode, failureInfo, testSuite)\;
                \If{rootCause is not NONE}{
                    Close CPG Project\;
                    isEnoughDonor $\leftarrow$ True\;
                }
            }
            \Return rootCause\;
        }
    }
    \Start{Solution Phase}{
        queryInput $\leftarrow$ concatenateInfo(bugCode, rootCause)\;
        repairPattern $\leftarrow$ repair\_pattern\_retrieval\_tool(queryInput)\;
        repairsSuggestions $\leftarrow$ giveRepairSuggestions(bugCode, rootCause, repairPattern)\;
        candidatePatches $\leftarrow$ patchGeneration(bugCode, repairSuggestions)\;
        pPatch $\leftarrow$ patchValidation(candidatePatches)\;
        \Return pPatch
        }
}
    \vspace{-0.05in}
\end{algorithm}

\begin{figure}[]
    \centering
        {\includegraphics[width=1\linewidth]{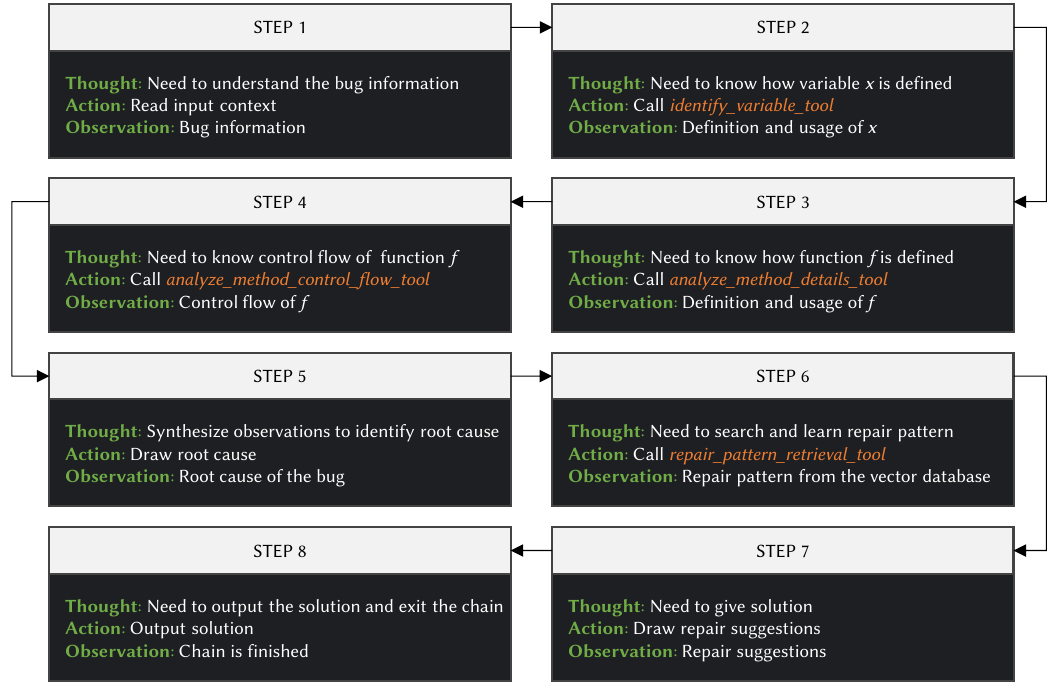}}
    \caption{An execution chain of the ReAct Agent.
    }
    \label{fig:react_chain}
\end{figure}

 Figure~\ref{fig:react_chain} illustrates the sequential Thought-Action-Observation cycle of the ReAct Agent, showing how it derives the root cause and repair suggestions for a bug. The agent reads the bug information and follows the Algorithm~\ref{alg:RepairIngredientFixer}'s analysis process to address complex tasks while avoiding potential infinite loops during reasoning and action. After gathering all relevant contextual code elements and learning from matched repair patterns, the agent provides a detailed explanation of the root cause, along with several repair suggestions as the Final Answer, which guides the subsequent patch generation process. For each root cause and suggested fix pair, we ask the model to generate multiple patches. These patches are then evaluated by running them against the test suite. Following previous works~\cite{ChatRepair, ThinkRepair, RepairAgent}, patches that pass all test cases are considered \textit{plausible patches}, while those verified by humans and deemed semantically equivalent to the ground truth are classified as \textit{correct patches}.



%% file: tables/Joern_Tools.tex

\begin{table}[t]
\centering
\caption{The overview of code analysis tools.}
\label{tab:joern-tools}
\vspace{-0.1in}
\resizebox{1\columnwidth}{!}{
\begin{tabular}{llll}
\toprule
\multicolumn{1}{c}{Tool Name} & \multicolumn{1}{c}{Type} & \multicolumn{1}{c}{Description}                                                             & \multicolumn{1}{c}{Usage Example}                   \\ \midrule
identify\_variable            & Variable-Level           & Analyzes variable definitions and usages in specific files                                  & \texttt{varName, fileName}    \\
find\_variable\_assignments   & Variable-Level           & Tracks all assignments to a specific variable                                               & \texttt{varName, fileName}    \\
track\_variable\_dataflow     & Variable-Level           & Analyzes data flow showing value sources and destinations                                   & \texttt{varName, fileName}    \\
trace\_method\_usage          & Method-Level            & Trace how and which file the method is used                                                 & \texttt{methodName}                \\
analyze\_method\_details      & Method-Level             & Comprehensive method analysis                                                               & \texttt{methodName, fileName} \\
find\_method\_in\_file        & Method-Level             & Analyzes the method control flow (if, while, for, etc.) structures & \texttt{methodName, fileName} \\
find\_class\_loc               & Class-Level              & Locate where the class is defined                            & \texttt{className}                 \\
identify\_class               & Class-Level              & Retrieves class definition code including attributes and methods                            & \texttt{className}                 \\

get\_imports                  & File-Level               & Retrieves all import statements from a specific file                                        & \texttt{fileName}      

\\ \bottomrule      
         
\end{tabular}
}
\vspace{-0.1in}
\end{table}

%% file: 4-setup.tex
\section{Experiment Setup}

\subsection{Research Questions}

\begin{itemize}[leftmargin=0.3cm]
    \item \textbf{RQ1: How does the repair effectiveness of \TOOL \space compare to state-of-the-art APR techniques?}
    This question evaluates \TOOL's repair capabilities in Java bug repair and compares its performance to existing baselines. (Repair Effectiveness)
    
    \item \textbf{RQ2: How does \TOOL \space perform across different repair scenarios?}
    This question investigates \TOOL's performance in three distinct repair scenarios~\cite{ChatRepair,ThinkRepair}, e.g., single-function, single-hunk, and single-line bugs. 
    (Repair Scenarios)

    \item \textbf{RQ3: How do different design choices in \TOOL \space impact its repair capabilities?}
    This question explores how \TOOL's novel repair ingredients search strategies affect the overall repair performance through an ablation study.
    (Ablation Study)

    \item \textbf{RQ4: Can \TOOL \space generalize to recent bugs post-LLM training data cut-off?}
    This question assesses \TOOL's generalization ability by testing its performance on recent real-world bugs,
    ensuring no data leakage. (Generalizability Study)

\end{itemize}

\input{tables/Defects4J}

\subsection{Benchmarks}
\begin{itemize}[leftmargin=0.3cm]
    \item \textbf{Defects4J.}
    To evaluate the repair effectiveness, we use the widely studied benchmark Defects4J~\cite{Defects4j}. Similar to prior APR studies~\cite{ChatRepair,ThinkRepair}, we separate Defects4J into V1.2 and V2.0. Defects4J V1.2 consists of 391 bugs (removed 4 deprecated bugs), and Defects4J V2.0 introduces 438 new bugs. In this work, we follow prior studies~\cite{ChatRepair,ThinkRepair} and categorize Defects4J into three repair scenarios, including single-function (SF), single-hunk (SH), and single-line (SL) bugs. Note that single-hunk is a subset of single-function and single-line is a subset of single-hunk bugs. 
    In addition, we will also implement fixes on the multi-function (MF) scenario.
    The statistic of each scenario is presented in Table~\ref{tab:details_d4j}.
    \kai{I removed this table to save page space, please add it on the camera ready version}
    \item \textbf{RWB.}
    To evaluate the generalizability study, we use the recent repair benchmark RWB~\cite{ThinkRepair}. Note that the pre-training data for ChatGPT (GPT-3.5~\cite{GPT4-3.5}) was collected before September 2021~\cite{GPT3_Training_Date}, and the pre-training data for DeepSeek-Coder was collected from GitHub before February 2023~\cite{deepseekcoder}. To assess data leakage of ChatGPT and DeepSeek, Yin \textit{et al.}~\cite{ThinkRepair} collected two datasets to assess data leakage. The first dataset (RWB V1.0) comprises bug-fixing commits after October 2021, while the second dataset (RWB V2.0) includes bug-fixing commits after March 2023, resulting 44 and 29 single-function bugs, respectively. 
\end{itemize}

\subsection{Baselines}

We considered recent APR works to serve as baselines. This includes:
ChatRepair~\cite{ChatRepair}, ThinkRepair~\cite{ThinkRepair}, RepairAgent~\cite{RepairAgent}, FitRepair~\cite{FitRepair}, GAMMA~\cite{GAMMA}, TENURE~\cite{TENURE}, Tare~\cite{Tare}, AlphaRepair~\cite{AlphaRepair}, RAP-Gen~\cite{RAP-Gen}, KNOD~\cite{KNOD}, Recoder-T~\cite{Recoder,Tare}, TBar~\cite{TBar}. Following the common practice in the APR community~\cite{ThinkRepair,ChatRepair,FitRepair,GAMMA,Tare,AlphaRepair,RAP-Gen,Recoder}, we reuse the reported results from previous studies~\cite{ThinkRepair,ChatRepair,FitRepair,GAMMA,Tare,AlphaRepair,RAP-Gen,TENURE,KNOD} instead of directly running the APR tools.

    

\subsection{Vector Database}
To construct a vector database for the external ingredients search, we utilize the TRANSFER dataset~\cite{TRANSFER} as the retrieval corpus and the text-embedding-3-large~\cite{Vector_Embeddings} as the embedding model.
TRANSFER dataset contains about one million BFPs and corresponding fix templates that have been used to drive the template-based APR works~\cite{TRANSFER,TENURE,NTR}. 
Considering the cost of embedding, we randomly selected 100K samples from TRANSFER for the vector database construction. We use the exact match strategy to filter out overlapping samples with those in benchmarks to avoid data leakage.

\subsection{Implementation}

\TOOL~is built on top of the LangChain~\cite{LangChain} framework, where we use gpt-3.5-turbo~\cite{GPT4-3.5} and gpt-4o~\cite{GPT4o} as base models, and set a sampling temperature of 1.
In fault localization (FL), to avoid additional biases introduced by the FL tool, we follow recent works~\cite{ChatRepair,ThinkRepair} under conditions of perfect fault localization. 
When generating patches, we limit the number of repair attempts to a maximum of 3 per bug. For each attempt, the maximum number of repair suggestions is set to 3, and for each repair suggestion, up to 5 candidate patches are considered. Therefore, the total maximum patch space size is 3*3*5=45.
For validation, only the top-1 plausible patch that passes all test cases is retained.
Finally, the plausible patches are manually reviewed to verify the semantic correctness. 

%% file: tables/Defects4J.tex
\begin{table}[h]
\vspace{-0.1in}
\caption{Statistics of Defects4J.}
\vspace{-0.1in}
\label{tab:details_d4j} 
\resizebox{1\columnwidth}{!}{
\begin{tabular}{c|cccc|cccc}
\toprule
\textbf{Benhmarks}        & \multicolumn{4}{c|}{\textbf{Defects4J V1.2}}                          & \multicolumn{4}{c}{\textbf{Defects4J V2.0}}                              \\
\textbf{Repair Scenarios} & \multicolumn{1}{l}{\# MF Bugs} & \# SF Bugs & \# SH Bugs & \# SL Bugs & \multicolumn{1}{l}{\# MF Bugs} & \# SF Bugs & \# SH Bugs & \# SL Bugs \\ \midrule
\textbf{\# Bug Num}       & 136                               & 255        & 154        & 80         & 210                               & 228        & 159        & 78        \\ \bottomrule
\end{tabular}
}
\vspace{-0.2in}
\end{table}

%% file: 5-evaluation.tex
\section{Evaluation}

\subsection{RQ1: Repair Effectiveness}

\input{tables/RQ1_D4J_V1}

\input{tables/RQ2_D4J_V1_and_2}

In the main experiment, we will investigate the overall repair effectiveness of \textsc{ReinFix}~on the Defects4J.
Table~\ref{tab:result_d4jv1} presents the repair results of the specific implementation of \textsc{ReinFix}. 

\textbf{Defects4J V1.2.}
As shown in Table~\ref{tab:result_d4jv1}, the implementations of \textsc{ReinFix}~outperform the current best baseline work, the ChatGPT-based APR tool ChatRepair~\cite{ChatRepair}. Specifically, the GPT3.5-based \textsc{ReinFix}$_{GPT3.5}$ fixes 4 more bugs (118 vs. 114) than ChatRepair, and the GPT4o-based \textsc{ReinFix}$_{GPT4o}$ fixed 32 more bugs (146 vs. 114) than ChatRepair. 
In particular, under the same condition of using GPT-3.5 as the foundation model, \textsc{ReinFix}~still outperforms the recent approaches ChatRepair and ThinkRepair, fixing 4 and 20 more bugs, respectively. This indicates that \textsc{ReinFix}'s strategic design not only provides a competitive edge over similar methods but also more effectively harnesses the repair potential of the LLM.
It is important to note that ThinkRepair is specifically designed for the single-function repair scenario, where its reported results include 98 successfully fixed single-function bugs. In contrast, the repair outcomes for \textsc{ReinFix}~and ChatRepair encompass both single-function and multi-function bugs. To provide a more direct comparison with the state-of-the-art single-function repair tool, ThinkRepair, we focus on the single-function repair scenario. In this context, \textsc{ReinFix}$_{GPT3.5}$ successfully repairs 104 single-function bugs, surpassing ThinkRepair by 6 cases.
Overall, \textsc{ReinFix}~not only outperforms the best-performing baseline, ChatRepair, in terms of overall repair effectiveness, but also performs on par with ThinkRepair in the single-function repair scenario.

\textbf{Defects4J V2.0.}
As shown in Table~\ref{tab:result_d4jv1}, \st{on the latest version of Defects4J benchmark,}\textsc{ReinFix}~ again demonstrates superior performance compared to the current leading baseline, the ChatGPT-based APR tool ThinkRepair~\cite{ChatRepair}.
Specifically, the GPT3.5-based \textsc{ReinFix}$_{GPT3.5}$ fixes 16 more bugs (123 vs. 107) than ThinkRepair, and the GPT4o-based \textsc{ReinFix}$_{GPT4o}$ fixed 38 more bugs (145 vs. 107) than ThinkRepair. 
As mentioned previously, unlike \textsc{ReinFix}, which aims to address all bug types, ThinkRepair~\cite{ThinkRepair} is specifically designed for single-function repairs, while ChatRepair~\cite{ChatRepair} and other baselines~\cite{FitRepair,AlphaRepair} focus only on single-line repair scenarios. To ensure a fair comparison with these recent GPT-3.5-based APR tools, we further analyze the repair capabilities of \textsc{ReinFix}$_{GPT3.5}$ within single-function and single-line repair scenarios.
In the single-function repair scenario, \textsc{ReinFix}$_{GPT3.5}$ successfully fixes 2 more bugs than ThinkRepair (109 vs. 107). 
In the single-line repair scenario, \textsc{ReinFix}$_{GPT3.5}$ also rivals ChatRepair's repair results (47 vs. 48). Overall, \textsc{ReinFix}~not only demonstrates an outstanding repair capability but also rivals or exceeds recent LLM-based APR methods in both single-function and single-line repair scenarios.

\begin{figure}
  \centering
  \vspace{-1px}
  {\includegraphics[width=1.0\linewidth, trim=240 109 240 120, clip]{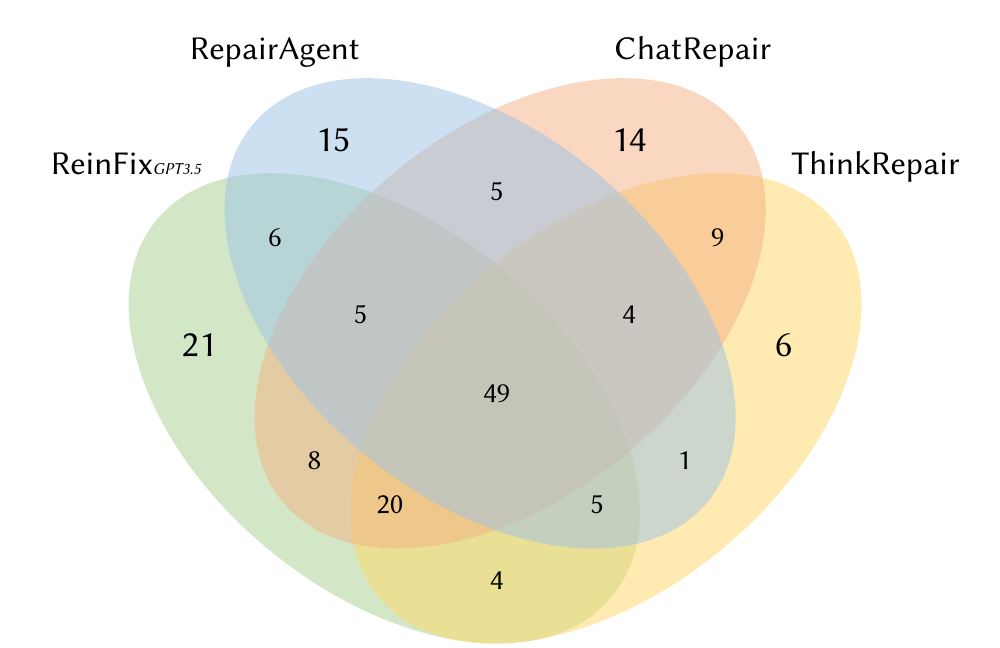}}
  \caption{Bug fix Venn diagram on Defects4J V1.2.}
  \label{fig:venn_result}
  \vspace{-10px}
\end{figure}

\textbf{Unique Fixes.}
Specifically, we also present ReinFix's unique repair capabilities compared to other APR tools. Specifically, we follow the practice of selecting Defects4J V1.2 in the baseline work~\cite{ChatRepair} to present unique fixes and keep the same base model GPT3.5 as recent LLM-based baselines~\cite{ChatRepair,ThinkRepair,RepairAgent}. As shown in Figure~\ref{fig:venn_result}, \textsc{ReinFix}$_{GPT3.5}$ still maintains an advantage in unique repair capabilities, achieving 21 unique fixes compared to recent APR tools using the same base model. These results indicate that \textsc{ReinFix}'s approach is complementary to existing work, emphasizing the necessity of searching for repair ingredients to enhance existing APR workflow.

In summary, \textsc{ReinFix} achieves the best bug repair performance on both Defects4J V1.2 and V2.0. In particular, \textsc{ReinFix}~maintains an advantage over recent approaches when using the same foundation model. These results highlight the effectiveness of \textsc{ReinFix}’s repair strategy and suggest that it represents a promising direction.

\subsection{RQ2: Repair Scenarios}
While the main experiments provide an overview of \textsc{ReinFix}'s overall repair effectiveness, examining its behavior within specific scenarios such as single-line, single-hunk, and single-function, allows us to assess its performance across different dimensions of complexity. 
Therefore, here we explore the multidimensional capabilities of \textsc{ReinFix} across various repair scenarios~\cite{ThinkRepair}.
By analyzing each repair scenario individually, following recent work~\cite{ThinkRepair,ChatRepair}, we aim to reveal \textsc{ReinFix}'s capabilities in greater detail and across multiple facets of the repair process.

As shown in Table~\ref{tab:result_three_repair_scenarios}, \textsc{ReinFix} achieves effective repair results across all tested scenarios, even without considering multi-function repair cases. Specifically, in the single-function repair scenario, \textsc{ReinFix}$_{GPT3.5}$ and \textsc{ReinFix}$_{GPT4o}$ repaired 6 and 26 more single-function bugs than the top-performing baseline, ThinkRepair~\cite{ThinkRepair}, on Defects4J V1.2, and 2 and 23 more single-function bugs, respectively, on Defects4J V2.0. 
In the single-hunk and single-line repair scenarios, \textsc{ReinFix}$_{GPT3.5}$ and \textsc{ReinFix}$_{GPT4o}$ continue to match or even exceed recent baseline performances. More importantly, we assessed \textsc{ReinFix} in the multi-function repair scenario, comparing it with RepairAgent~\cite{RepairAgent}, another tool supporting multi-function fixes. Here, \textsc{ReinFix}$_{GPT3.5}$ and \textsc{ReinFix}$_{GPT4o}$ again demonstrated higher fix rates.
Overall, these findings indicate that \textsc{ReinFix} exhibits robust performance across diverse repair scenarios.


\subsection{RQ3: Ablation Study}
\label{sec:RQ3}

In the ablation experiment, we aim to evaluate the contributions of \textsc{ReinFix}'s key components in enhancing the repair capabilities of LLMs and to analyze the advantages of our repair ingredient retrieval methods (i.e., Section~\ref{sec_approach:donor} and Section~\ref{sec_approach:pattern}) in comparison to previous work.
Therefore, we designed several variants to reveal the effects of different design choices. Specifically, we created six \textsc{ReinFix} variants:
\ding{182} \textsc{ReinFix}$_{NN}$: We removed two key components from the original \textsc{ReinFix}~implementation to establish a basic baseline that represents the fundamental repair capabilities of ChatGPT.
\ding{183} \textsc{ReinFix}$_{DN}$: In this variant, we investigate the impact of \textsc{ReinFix}'s Dependency-Analysis-Based internal ingredient search approach (Section~\ref{sec_approach:donor}) on repair effectiveness. Only the internal ingredient search component is used for contextual code search, assisting LLMs in better analyzing the root cause of the bugs.
\ding{184} \textsc{ReinFix}$_{NP}$: This variant explores the impact of \textsc{ReinFix}'s Pattern-Matching-Based external ingredients search approach (Section~\ref{sec_approach:pattern}) on repair effectiveness. Only the external ingredient search component is used for bug pattern retrieval, assisting LLMs in better identifying repair behaviors to generate patches.
\ding{185} \textsc{ReinFix}$_{FP}$: This variant explores the advantages of \textsc{ReinFix}'s dependency-analysis-based internal ingredient search approach over the search schemes used in recent work~\cite{FitRepair}. The original internal ingredient search component of \textsc{ReinFix} is replaced with FitRepair's repair ingredient retrieval scheme to observe the impact of using different contextual code search strategies on the repair results.
\ding{186} \textsc{ReinFix}$_{DR}$: To explore the advantages of \textsc{ReinFix}'s pattern-matching-based external ingredient search approach over the retrieval schemes used in recent work~\cite{RAP-Gen}, we replaced \textsc{ReinFix}'s original external ingredients search component with RAP-Gen's repair ingredient retrieval scheme. This allows us to observe the impact of using different search strategies on the repair results.
\ding{187} \textsc{ReinFix}$_{DP}$: For reference, we include the original implementation of \textsc{ReinFix}~with both key repair ingredients search components, showcasing the full capabilities of the system.

In this ablation study, we select GPT-4o as the foundational model and Defects4J V1.2 as the benchmark for testing. We will now analyze the contributions and advantages of the design choices in \textsc{ReinFix} by evaluating several variants of the system.

\input{tables/RQ3_D4J_V1}

\subsubsection{Contributions of \textsc{ReinFix} in Design Choices}
By comparing the repair results of the \textsc{ReinFix} variants (\ding{182}, \ding{183}, \ding{184}, and \ding{187}), we can examine the contributions of \textsc{ReinFix}'s key components and evaluate their roles in enhancing the repair capabilities of LLMs.

\textbf{1) \ding{182} \textsc{ReinFix}$_{NN}$ vs. \ding{183} \textsc{ReinFix}$_{DN}$:}
The first key component of the \textsc{ReinFix}~is to help LLM reason about possible bug root causes by internal ingredient search. By comparing the repair results of \textsc{ReinFix}$_{NN}$ and \textsc{ReinFix}$_{DN}$, we can directly observe the impact of including the first key component on the repair capability of LLMs. As shown in Table~\ref{tab:ab_RQ3}, \textsc{ReinFix}$_{DN}$ fixes 31 more bugs (116 compared to 85) than \textsc{ReinFix}$_{NN}$. This result indicates that the first key component is crucial for enhancing the repair capability of LLMs. By reasoning about the root cause of bugs and retrieving relevant internal donor code through the dependency-analysis based approach, the LLM gains a more comprehensive understanding of the bugs, which enables it to generate more accurate fix patches and achieve better repair results.

\textbf{2) \ding{182} \textsc{ReinFix}$_{NN}$ vs. \ding{184} \textsc{ReinFix}$_{NP}$:}
The second key component of the \textsc{ReinFix}~is to help LLM generate correct repair actions by external ingredient search. By comparing the repair results of \textsc{ReinFix}$_{NN}$ and \textsc{ReinFix}$_{NP}$, we can directly observe the impact of including the second key component on enhancing the repair capability of LLMs.
Since \textsc{ReinFix}$_{NP}$ removes the first key component, the LLM instead independently reasons about the root cause of the bug to complete the fix.
As shown in Table~\ref{tab:ab_RQ3}, \textsc{ReinFix}$_{NP}$ fixes 23 more bugs (108-85) than \textsc{ReinFix}$_{NN}$, indicating that the second key component is also crucial for enhancing the repair capability of LLMs. By generating correct repair patches through the search for relevant repair behaviors, the LLM can enrich its fix-related domain knowledge, leading to improved repair results.

\textbf{3) \ding{187} \textsc{ReinFix}$_{DP}$ vs. \ding{182} \textsc{ReinFix}$_{NN}$ \ding{183} \textsc{ReinFix}$_{DN}$ \ding{184} \textsc{ReinFix}$_{NP}$:}
It is important to note that the complete implementation of \textsc{ReinFix}~is a two-phase workflow. In the root cause analysis (i.e., reasoning) phase, it searches for internal repair ingredients to help better understand the root cause of the bug. In the repair pattern matching (i.e., solution) phase, it uses the root cause obtained in the previous phase to help search for relevant repair actions and implement patch generation. Indeed, there is a dependency between the two key components of \textsc{ReinFix}~and their synergies also deserve attention.
As shown in Table~\ref{tab:ab_RQ3}, \textsc{ReinFix}$_{DP}$ not only fixes 61 more bugs than \textsc{ReinFix}$_{NN}$ but also fixes 30 and 38 more bugs than \textsc{ReinFix}$_{DN}$ and \textsc{ReinFix}$_{NP}$, respectively. These results demonstrate that both key components of \textsc{ReinFix}~are crucial for enhancing the base repair capability of LLMs, and using them together maximizes the overall repair effectiveness.

\begin{figure}[]
    \centering
        {\includegraphics[width=1\linewidth]{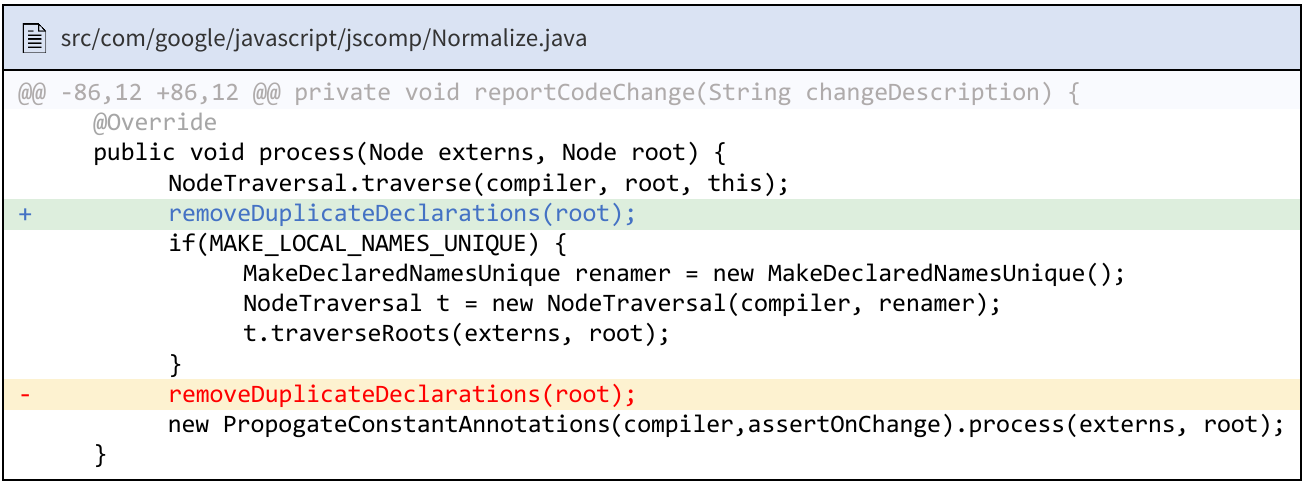}}
    \caption{The human patch of Closure-102.}
    \label{fig:closure-102}
\end{figure}


\begin{figure}
\centering
\vspace{-0.1in}
\subfloat[Retrieved repair ingredients of \textsc{ReinFix}$_{DP}$.]{
		\includegraphics[width=1\linewidth]{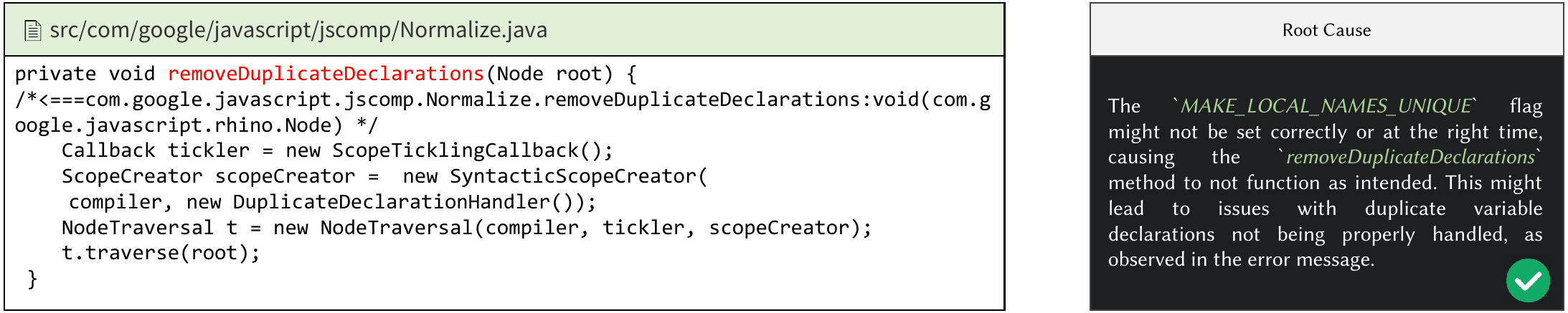}} \\
\subfloat[Retrieved repair ingredients of \textsc{ReinFix}$_{FP}$.]{
		\includegraphics[width=1\linewidth]{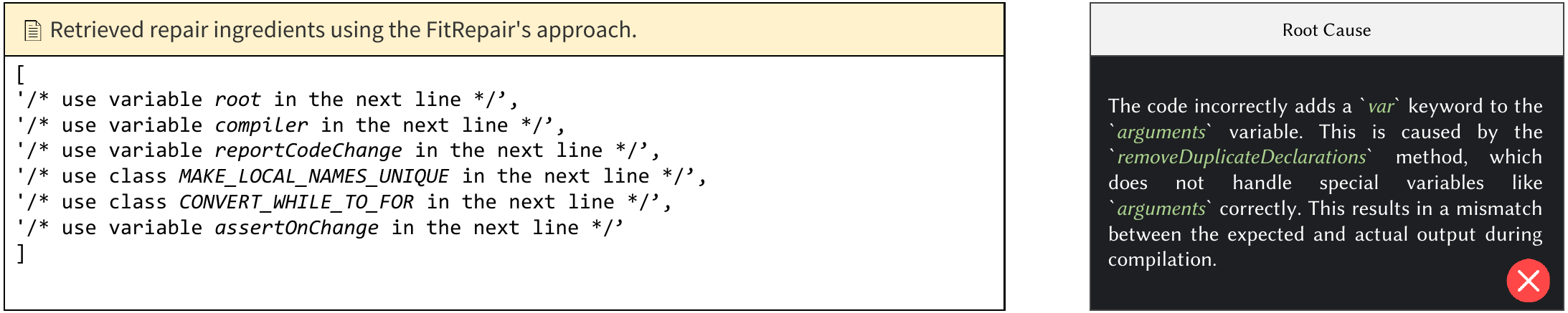}}
\caption{The Repair process of \textsc{ReinFix}$_{DP}$ and \textsc{ReinFix}$_{FP}$ for Closure-102.
}
\label{fig:closure-102-case}
\end{figure}

\input{tables/RQ4_RWB}

\subsubsection{Advantages of \textsc{ReinFix} in Design Choices.}
By comparing the results of \textsc{ReinFix}~variants \ding{185} \ding{186} \ding{187}, we can reveal the advantages of our repair ingredients search methods over previous works~\cite{FitRepair,RAP-Gen}.

\textbf{1) \ding{187} \textsc{ReinFix}$_{DP}$ vs. \ding{185} \textsc{ReinFix}$_{FP}$:}
In the first key component design of \textsc{ReinFix}, we adopted a program-dependency-based approach to search for internal donor code repair ingredients to assist in root cause analysis. A closely related work, FitRepair~\cite{FitRepair}, employs a code-similarity-based approach~\cite{Levenshtein_Distance} to retrieve relevant identifiers for prompt augmentation. While \textsc{ReinFix}~and FitRepair differ in their respective purposes for leveraging repair ingredients, they also adopt distinct design choices for searching these ingredients. To fairly evaluate the advantages of \textsc{ReinFix}'s design choices over those of FitRepair, we re-implemented the first key component of \textsc{ReinFix}~using FitRepair's similarity-based approach for direct comparison.
As shown in Table~\ref{tab:ab_RQ3}, \textsc{ReinFix}$_{DP}$ fixed 36 more bugs than \textsc{ReinFix}$_{FP}$, demonstrating the superiority of \textsc{ReinFix}'s donor code search strategy over FitRepair's approach. Upon analysis, we found that repair ingredient retrieval methods that do not consider program dependencies often introduce irrelevant or incorrect donor code. This can mislead the model into inferring incorrect root causes and result in repair failures. By leveraging program dependencies, \textsc{ReinFix} ensures the retrieval of contextually accurate ingredients, which is critical for effective bug fixing.
Here, we observe the case, Closure-102 (Figure~\ref{fig:closure-102}), where \textsc{ReinFix}$_{DP}$ successfully repaired the bug, while \textsc{ReinFix}$_{FP}$ failed. 
As shown in Figure~\ref{fig:closure-102-case}-(a), \textsc{ReinFix}$_{DP}$ accurately identifies the definition of the buggy method during the reasoning phase by leveraging program dependency analysis, enabling the model to infer the correct root cause. Conversely, as illustrated in Figure~\ref{fig:closure-102-case}-(b), \textsc{ReinFix}$_{FP}$ retrieves repair ingredients based solely on code similarity, resulting in irrelevant donor code. These irrelevant ingredients fail to assist the model in analyzing the critical root cause, causing it to produce incorrect fixes.

\textbf{2) \ding{187} \textsc{ReinFix}$_{DP}$ vs. \ding{186} \textsc{ReinFix}$_{DR}$:}
In the second key component design of \textsc{ReinFix}, we adopted a pattern matching-based approach to search external repair ingredients for similar repair patterns. The closest work is RAP-Gen~\cite{RAP-Gen}, which also follows the RAG paradigm to retrieve relevant fix patterns for prompt augmentation. The key difference is that RAP-Gen's method does not consider high-level bug root cause information. To fairly evaluate the advantages of \textsc{ReinFix}'s design choices over RAP-Gen, we reimplemented the second key component of \textsc{ReinFix}~using RAP-Gen's retrieval approach for direct comparison.
As shown in Table~\ref{tab:ab_RQ3}, \textsc{ReinFix}$_{DP}$ fixed 27 more bugs than \textsc{ReinFix}$_{DR}$, indicating that \textsc{ReinFix}'s  search strategy outperforms RAP-Gen's approach. We analyzed the reasons behind it and found that relying solely on source code-level (the function) similarity is insufficient, as similar code may have different bugs. In contrast, by incorporating bug root cause information, \textsc{ReinFix} effectively guides the retrieval of relevant repair actions. This highlights the importance of considering the root cause in repair pattern retrieval, as bugs with similar root causes are likely to require similar repair behaviors.
For example, in Closure-51 (Figure~\ref{fig:closure-51}), \textsc{ReinFix}$_{DP}$ retrieved a relevant repair behavior \texttt{(value!=0 || !negative)}, whereas \textsc{ReinFix}$_{DR}$ retrieved an irrelevant repair behavior \texttt{if (pos-start >= 10)}. 
This highlights that high-level root cause is critical not only for understanding the bug but also for retrieving accurate repair behaviors.

\subsection{RQ4: Generalizability Study}
\label{sec:RQ4}

In the generalizability experiments, we focus on the performance of the \textsc{ReinFix} on additional benchmarks as well as other foundation models. Here, we follow ThinkRepair~\cite{ThinkRepair} to test on the RWB benchmark and implement \textsc{ReinFix}$_{DSC}$ with DeepSeek-Coder~\cite{deepseekcoder} as the foundation model. Note that the RWB V1.0 and RWB V2.0 collect bug cases after the training cutoff date for GPT-3.5 and DeepSeek-Coder, respectively, and thus the RWB dataset can further mitigate the data leakage risks of LLMs during the repair process.
As shown in Table~\ref{tab:result_rwb}, \textsc{ReinFix}$_{GPT3.5}$ fixes 1 more bug than ThinkRepair on RWB V1.0, and \textsc{ReinFix}$_{DSC}$ fixes 2 more bugs than ThinkRepair* on RWB V2.0. Overall, \textsc{ReinFix}~achieves a 10.34\% improvement over ThinkRepair.
These results demonstrate that even when minimizing the risk of data leakage and using the same base model, \textsc{ReinFix}~remains effective and outperforms the best baseline.

Additionally, we tested several \textsc{ReinFix} implementations using other LLMs to explore whether our framework remains effective across different models. Specifically, considering the training data cutoff for RWB V1.0 and V2.0, we selected GPT-4 (i.e., gpt-4-0613~\cite{GPT4}) and GPT-4-turbo (i.e., gpt-4-1106-preview~\cite{GPT4-turbo}) to evaluate the framework on these two benchmark versions, respectively. As shown in Table~\ref{tab:result_rwb}, \textsc{ReinFix}$_{GPT4}$ and \textsc{ReinFix}$_{GPT4t}$ fixed 21 and 15 bugs on RWB V1.0 and V2.0, respectively, achieving the best repair results. These results highlight that the \textsc{ReinFix} framework demonstrates strong generalization, effectively producing fixes not only on additional benchmarks but also across a range of LLMs.


%% file: tables/RQ1_D4J_V1.tex
\begin{table*}[]
\caption{Repair results (correct fixes / plausible fixes) for \textsc{ReinFix} and baselines on Defects4J V1.2 and V2.0. 
}
\label{tab:result_d4jv1} 
\resizebox{2.1\columnwidth}{!}{
\begin{tabular}{ccccccccccccccc}
\toprule
\textbf{APR Tool}     & \textbf{\textsc{ReinFix}$_{GPT4o}$} & \textbf{\textsc{ReinFix}$_{GPT3.5}$} & \textbf{ChatRepair} & \textbf{ThinkRepair} & \textbf{RepairAgent} &\textbf{FitRepair} & \textbf{GAMMA} & \textbf{TENURE} & \textbf{Tare} & \textbf{AlphaRepair} & \textbf{RAP-Gen} & \textbf{KNOD} & \textbf{Recoder} & \textbf{TBar} \\
Sampling Times        & 3*3*5          & 3*3*5           & 500                 & 25*5          & 117        & 1000*4             & 250            & 500             & 100           & 5000                 & 100              & 1000          & 100              & -             \\ \midrule
Chart                 & 18/20           & 16/17            & 15/-                & 11/-           & 11/14      & 8/-                & 11/11          & 7/-             & 11/-          & 9/-                  & 9/-              & 10/11         & 9/-              & 11/-          \\
Closure               & 40/50           & 30/37            & 37/-                & 31/-        & 25/25         & 29/-               & 24/26          & 26/-            & 25/-          & 23/-                 & 22/-             & 23/29         & 25/-             & 16/-          \\
Lang                  & 33/47           & 26/33            & 21/-                & 19/-        & 17/17         & 19/-               & 16/25          & 16/-            & 14/-          & 13/-                 & 12/-             & 11/13         & 12/-             & 13/-          \\
Math                  & 39/68           & 35/52            & 32/-                & 27/-        & 29/29         & 24/-               & 25/31          & 22/-            & 22/-          & 21/-                 & 26/-             & 20/25         & 20/-             & 22/-          \\
Mockito               & 10/11           & 8/9            & 6/-                 & 6/-           & 6/6       & 6/-                & 3/3            & 4/-             & 2/-           & 5/-                  & 2/-              & 5/5           & 2/-              & 3/-           \\
Time                  & 6/11           & 3/4            & 3/-                 & 4/-           & 2/3       & 3/-                & 3/5            & 4/-             & 3/-           & 3/-                  & 1/-              & 2/2           & 3/-              & 3/-           \\ \midrule
\#Total (D4J V1.2) & \textbf{146/207}        & \textbf{118/152}           & 114/-               & 98/-        & 90/94         & 89/-               & 82/101         & 79/-            & 77/-          & 74/109               & 72/-             & 71/85         & 71/-             & 68/95     \\ \midrule
\#Total (D4J V2.0) & \textbf{145/190}        & \textbf{123/147}           & 48/-               & 107/-       & 74/92          & 44/-               & 45/-         & 50/-            & -/-          & 36/-               & 53/-             & 50/85         & 11/-             & 8/-     \\
\bottomrule   
\end{tabular}
}
\end{table*}

%% file: tables/RQ2_D4J_V1_and_2.tex

\begin{table}[t]
\caption{Repair results (correct fixes) of different repair scenarios for \textsc{ReinFix} and baselines on Defects4J V1.2 and V2.0.}
\label{tab:result_three_repair_scenarios} 
\resizebox{1.0\columnwidth}{!}{
\begin{tabular}{c|cccc|cccc}
\toprule
\textbf{Benhmarks} & \multicolumn{4}{c|}{\textbf{Defects4J V1.2}}                                                              & \multicolumn{4}{c}{\textbf{Defects4J V2.0}}                                                              \\
Repair Scenarios   & \multicolumn{1}{l}{\cellcolor[HTML]{EFEFEF}\# MF Bugs} & \# SF Bugs & \# SH Bugs & \# SL Bugs & \multicolumn{1}{l}{\cellcolor[HTML]{EFEFEF}\# MF Bugs} & \# SF Bugs & \# SH Bugs & \# SL Bugs \\ \midrule
ChatRepair         & \cellcolor[HTML]{EFEFEF}-                                  & 76              & -           & -           & \cellcolor[HTML]{EFEFEF}-                                  & -               & -           & 48          \\
ThinkRepair        & \cellcolor[HTML]{EFEFEF}-                                  & 98              & 78          & 52          & \cellcolor[HTML]{EFEFEF}-                                  & 107             & 81          & 47          \\
RepairAgent        & \cellcolor[HTML]{EFEFEF}7                                  & 83               & 71           & 51           & \cellcolor[HTML]{EFEFEF}6                                  & 68               & 65           & 48           \\
\textsc{ReinFix}$_{GPT3.5}$              & \cellcolor[HTML]{EFEFEF}14                                  & 104             & 78          & 53          & \cellcolor[HTML]{EFEFEF}14                                  & 109             & 85          & 47          \\
\textsc{ReinFix}$_{GPT4o}$               & \cellcolor[HTML]{EFEFEF}22                                  & 124             & 93          & 57          & \cellcolor[HTML]{EFEFEF}15                                  & 130             & 103         & 56       \\ \bottomrule  
\end{tabular}
}
\vspace{-0.1in}
\end{table}

%% file: tables/RQ3_D4J_V1.tex
\begin{table}[t]
\caption{Repair results (correct fixes) of \textsc{ReinFix} when using different ingredients search components on Defects4J V1.2.
}
\label{tab:ab_RQ3} 
\resizebox{1.0\columnwidth}{!}{
\begin{tabular}{c|cc|c}
\toprule
\multirow{2}{*}{\textbf{Variants}} & \multicolumn{2}{c|}{\textbf{Repair Ingredients Search Components}}                                  & \multirow{2}{*}{\textbf{Repair Result}} \\
                                   & \multicolumn{1}{c}{Internal Ingredients Search} & \multicolumn{1}{c|}{External Ingredients Search} &                                         \\ \midrule
\ding{182} \textsc{ReinFix}$_{NN}$                              &NULL                                          &NULL                                             & 85                                  \\ \midrule
\ding{183} \textsc{ReinFix}$_{DN}$                              & Dependency-Analysis-Based     &NULL                                             & 116                                 \\ \midrule
\ding{184} \textsc{ReinFix}$_{NP}$                              &NULL                                          & Pattern-Matching-Based         & 108                                 \\ \midrule
\ding{185} \textsc{ReinFix}$_{FP}$                               & FitRepair~\cite{FitRepair}                                      & Pattern-Matching-Based         & 110                                     \\ \midrule
\ding{186} \textsc{ReinFix}$_{DR}$                              & Dependency-Analysis-Based     & RAP-Gen~\cite{RAP-Gen}                                           & 119                                 \\ \midrule
\ding{187} \textsc{ReinFix}$_{DP}$                             & Dependency-Analysis-Based     & Pattern-Matching-Based         & 146        \\ \bottomrule                        
\end{tabular}
}
\vspace{-0.1in}
\end{table}

%% file: tables/RQ4_RWB.tex
\begin{table}[]
\caption{Repair results (correct fixes) for \TOOL~and baselines on RWB V1.0 (44 cases) and V2.0 (29 cases).
}
\label{tab:result_rwb} 
\resizebox{1\columnwidth}{!}{
\begin{tabular}{c|cccc|cccc}
\toprule
\textbf{Benchmark}                      & \multicolumn{4}{c|}{\textbf{RWB V1.0 (44 bugs)}}                & \multicolumn{4}{c}{\textbf{RWB V2.0 (29 bugs)}}                        \\  
\textbf{APR Tool}                 & \textbf{\textsc{ReinFix}$_{GPT4}$} & \textbf{\textsc{ReinFix}$_{GPT3.5}$}     & \textbf{ThinkRepair} & \textbf{AlphaRepair} & \textbf{\textsc{ReinFix}$_{GPT4t}$} & \textbf{\textsc{ReinFix}$_{DSC}$}           & \textbf{ThinkRepair*}  & \textbf{AlphaRepair} \\
\textbf{LLM}           & \textbf{GPT-4} & \textbf{GPT-3.5} & \textbf{GPT-3.5}     & \textbf{CodeBERT}    & \textbf{GPT-4-turbo} & \textbf{DeepSeek} & \textbf{DeepSeek} & \textbf{CodeBERT}    \\ \midrule
Cli                               & 4 & 4                  & 4                    & 3                    & 4 & 4                       & 4                      & 3                    \\
Codec                             & 3 & 3                 & 3                    & 1                    & 3 & 2                       & 1                      & 1                    \\
Collections                       & 1 & 1                 & 1                    & 0                    & - & -                       & -                      & -                    \\
Compress                          & 2 & 2                 & 1                    & 1                    & 1 & 1                       & -                      & -                    \\
Csv                               & 1 & 1                 & 1                    & 0                    & - & -                       & -                      & -                    \\
Jsoup                             & 7 & 6                 & 6                    & 2                    & 4 & 2                       & 2                      & 1                    \\
Lang                              & 3 & 3                 & 3                    & 2                    & 3 & 3                       & 3                      & 1                    \\
 \midrule
\textbf{\# Total} & \textbf{21} & \textbf{20}                 & 19                   & 9                    & \textbf{15} & \textbf{12}                       & 10                     & 6         \\ \bottomrule          
\end{tabular}
}
\end{table}


%% file: 6-threats.tex
\section{Threats to Validity}


\textbf{Data Leakage.}
The training data of LLMs may contain correct patches for buggy projects, potentially affecting the evaluation. 
To mitigate this risk, we follow previous work~\cite{ThinkRepair} and evaluate on bug cases collected after the cutoff date for LLM training data to ensure that the LLMs have not encountered the bug cases in the testing benchmarks during training. As shown in the generalizability experiments (Section~\ref{sec:RQ4}), \TOOL \space performs well on the recently collected RWB benchmarks, which helps minimize the impact of data leakage on the results. Moreover, our ablation experiments (Section~\ref{sec:RQ3}) demonstrate that the design choices in \TOOL \space further enhance the repair capability of base models, ensuring that the observed gains are independent of data leakage.

\noindent
\textbf{Repair Costs.}
Using LLMs may
lead to higher repair costs. To address this concern, we follow prior work~\cite{ChatRepair} that evaluates the average cost per bug fix. In our main experiment, \textsc{ReinFix}$_{GPT3.5}$, which uses GPT-3.5 as the base model, incurs an average cost of \$0.06 per bug fix, while \textsc{ReinFix}$_{GPT4o}$ with GPT-4o as the base model averages \$1.45 per bug fix. In comparison, ChatRepair, using GPT-3.5 as the base model, has an average cost of \$0.42 per bug fix. Thus, \textsc{ReinFix}$_{GPT3.5}$ not only reduces costs but also improves repair capability. We anticipate that this threat will be further reduced as LLM costs decrease with ongoing advancements in model efficiency.


%% file: 7-relatedwork.tex
\section{Related Work}
\chen{Look too short, as this is a very hot area, maybe better to split it into convetional ingredient related program repair and also LLM based repair.}
\kai{Given that the page limitation, we have to give a short intro about APR techniques.}
APR research has entered the era of LLMs~\cite{LLM4APR_Survey_Zhang}.
Researchers are devoted to designing novel repair strategies to maximize the repair potential of LLM. 
In the early LLM4APR studies,
researchers adopted the fine-tuning paradigm~\cite{LLM4APR_Study_Huang,LLM4APR_Study_Wu,LLM4APR_Study_Jiang} to allow LLMs to learn empirical knowledge used to guide bug repair from massive bug-fix history. For example, work such as VulRepair~\cite{VulRepair}, CIRCLE~\cite{Circle}, and NTR~\cite{NTR} follow the fine-tuning paradigm to learn defect repair knowledge. In addition, researchers have also utilized the prompt-learning paradigm~\cite{LLM4APR_Study_Xia,AlphaRepair,GAMMA} to stimulate the repair potential of LLMs by combining LLMs with traditional APR techniques to construct specific prompt templates. For example, AlphaRepair~\cite{AlphaRepair} and GAMMA~\cite{GAMMA} generate patches directly for fault locations by leveraging traditional template-based techniques to design specific repair template prompts.
Later, with the rapid development of generative models, LLMs possess more powerful language comprehension and generation capabilities, and can receive longer context window sizes to process more complex information. 
Therefore, researchers proposed many ChatGPT-based APR tools. For example, ChatRepair~\cite{ChatRepair} introduces the conversational repair paradigm to make LLMs work better guided by feedback information such as test execution; SRepair~\cite{SRepair} leverages LLMs' programming language understanding to infer possible root causes to generate patches; ThinkRepair~\cite{ThinkRepair} enables LLMs to better implement reasoning to complete the repair by leveraging the chain of thought.
Recently, LLM4APR research has begun to enter the era of agent~\cite{SEAgent_Michael}, where researchers have allowed LLMs to take on different task roles,
further expanding the repair capabilities of LLMs. For example, RepairAgent~\cite{RepairAgent} enables the LLM-based agent to perform repair by invoking external tools, and FixAgent~\cite{FixAgent} realizes unified debugging by constructing the multi-agent collaboration paradigm.
Unlike previous works, our work effectively combines repair ingredients with LLM that searches for internal ingredients to better analyze bug causes, and searches for external ingredients to better guide bug fixing.

%% file: 8-conclusion.tex
\section{Conclusion}
In this work, we presented \TOOL, a framework that integrates repair ingredients to enhance the capabilities of LLMs for APR. \TOOL \space is designed as a two-phase approach including reasoning and solution. 
During the reasoning phase, we utilize dependency analysis for internal ingredients search to facilitate LLMs in root cause analysis. 
During the solution phase, we combine the buggy code and root cause as bug patterns for external ingredient search to guide the patch generation. 
The synergy of the two phases that build on the ReAct Agent improves the LLM’s ability to generate 
patches.
Our extensive experiments, 
including comparisons with state-of-the-art baselines and ablation studies, 
demonstrate that \TOOL \space consistently outperforms existing methods.